\documentclass[12pt]{article}

\usepackage{amssymb}
\usepackage{latexsym,amsmath,amsthm,amsfonts}
\usepackage{cite} 
\usepackage{url}  
\usepackage[round]{natbib}
\usepackage{graphicx}
\usepackage{authblk}
\usepackage{color}
\usepackage{subfig}

\newcommand{\G}{\mathbf{G}}

\newcommand{\bu}{\mathbf{u}}
\newcommand{\bh}{\mathbf{h}}
\newcommand{\bv}{\mathbf{v}}

\newcommand{\br}{\mathbf{r}}

\newcommand{\bG}{\mathbf{G}}
\newcommand{\bX}{\mathbf{X}}
\newcommand{\bx}{{\bf x}}

\newcommand{\0}{{\bf 0}}

\newcommand{\bp}{{\bf p}}
\newcommand{\bz}{{\bf z}}
\newcommand{\bZ}{{\bf Z}}
\newcommand{\brho}{\boldsymbol{\rho}}

\newcommand{\bzeta}{\boldsymbol{\zeta}}
\newcommand{\bgam}{\boldsymbol{\gamma}}
\newcommand{\bGam}{\boldsymbol{\Gamma}}
\newcommand{\balpha}{\boldsymbol{\alpha}}
\newcommand{\bmu}{\boldsymbol{\mu}}

\newcommand{\real}{\mathbb{R}}
\newcommand{\pr}{\mathbb{P}}
\newcommand{\In}{\mathbb{I}}
\newcommand{\E}{\mathbb{E}}

\DeclareMathOperator*{\argmax}{arg\,max}

\newcommand{\be}{\begin{equation}}
\newcommand{\ee}{\end{equation}}

\newcommand{\bes}{\begin{equation*}}
\newcommand{\ees}{\end{equation*}}

\newcommand{\bea}{\begin{eqnarray}}
\newcommand{\eea}{\end{eqnarray}}

\newtheorem{thm}{Theorem}[section]
\newtheorem{lem}[thm]{Lemma}

\newtheorem{prop}[thm]{Proposition}

\title{An Empirical Bayes Approach for Multiple Tissue eQTL Analysis}
\author[1]{Gen Li}
\author[2]{Andrey A.\ Shabalin}
\author[3]{Ivan Rusyn}
\author[4]{Fred A.\ Wright}
\author[5]{Andrew B.\ Nobel}

\affil[1]{Department of Biostatistics, Mailman School of Public Health, Columbia University}
\affil[2]{Center for Biomarker Research and Personalized Medicine, Virginia Commonwealth University}
\affil[3]{Texas Veterinary Medical Center, Texas A$\&$M University}
\affil[4]{Department of Statistics and Biological Sciences,North Carolina State University}
\affil[5]{Department of Statistics and Operations Research, University of North Carolina at Chapel Hill}

\begin{document}
\maketitle
\newpage

\begin{abstract}
{Expression quantitative trait locus (eQTL) analyses identify genetic markers associated with the expression of a gene. Most up-to-date eQTL studies consider the connection between genetic variation and expression in a single tissue. Multi-tissue analyses have the potential to improve findings in a single tissue, and elucidate the genotypic basis of differences between tissues. In this paper we develop a hierarchical Bayesian model (MT-eQTL) for multi-tissue eQTL analysis. MT-eQTL explicitly captures patterns of variation in the presence or absence of eQTL, as well as the heterogeneity of effect sizes across tissues. We devise an efficient Expectation-Maximization (EM) algorithm for model fitting. Inferences concerning eQTL detection and the configuration of eQTL across tissues are derived from the adaptive thresholding of local false discovery rates, and maximum a-posteriori estimation, respectively. We also provide theoretical justification of the adaptive procedure. We investigate the MT-eQTL model through an extensive analysis of a 9-tissue data set from the GTEx initiative.}

\end{abstract}

\newpage

\section{Introduction}

Genetic variation in a population is commonly studied through the analysis of single nucleotide polymorphisms
(SNPs), which are genetic variants occurring at specific sites in the genome.
Expression quantitative trait locus
(eQTL) analysis seeks to identify genetic variants affecting the expression of one or more genes:
a gene-SNP pair for which the expression of the gene is associated with the value of the SNP is
referred to as an eQTL.
Identification of eQTL has proven to be
a useful tool in the study of pathways and networks
that underlie disease in human and other populations \citep[cf.][]{kendziorski2006review,wright2012computational}.

To date, most eQTL studies have considered the effects of genetic variation on expression within a single tissue.
A natural next step in understanding the genomic variation of expression
is the simultaneous analysis of eQTL in multiple tissues.
Multi-tissue eQTL analysis has the potential to improve the findings of single tissue analyses
by borrowing strength across tissues, and to address fundamental biological questions about the nature and source of variation
between tissues.
An important feature of multiple tissue studies is that a SNP may be associated with
the expression of a gene in some tissues,
but not in others.  Thus a full multi-tissue analysis must identify complex patterns of association
across multiple tissues.

Until recently, understanding of multi-tissue eQTL relationships was limited by a shortage of true
multi-tissue data sets, requiring the assimilation of data or results from different studies involving distinct populations, measurement platforms, and analysis protocols. 
By contrast, the GTEx initiative \citep{ardlie2015genotype} and related projects are currently
generating genetic data from dozens of tissues in several hundred individuals, greatly expanding our
potential understanding
of eQTLs across multiple tissues.
The size and complexity of these emerging multi-tissue data sets have created the
need to expand existing statistical tools for eQTL analysis.

{\color{black} In this paper we introduce and study a hierarchical Bayesian model for the
simultaneous analysis of eQTL in multiple tissues.
We particularly focus on cis-eQTL, where a SNP is located near the transcription start site of a gene.
We call the method {\em MT-eQTL} (MT stands for multi-tissue).
The {\em dimension} of the MT-eQTL model is equal to the number of tissues.
In this paper, we primarily consider a moderate dimension, typically between 1 and 10.}
Importantly, we do not seek to model the full expression and genotype data, but focus
instead on the vector $\bz$ of Fisher transformed correlations between expression and
genotype across tissues.
Figure \ref{fig:right} (upper panel) shows a density
scatter plot of the z-statistics for the lung and thyroid data from GTEx pilot data freeze as reported by \cite{ardlie2015genotype}.
The lower panel illustrates the results of the MT-eQTL model: z pairs close to the origin for which no eQTL are
detected have been removed, resulting in the central white region; detected eQTL are colored
according to whether an eQTL is detected in both tissues (light gray points) or a single tissue
(dark gray and black points).
{\color{black}Our model explicitly captures patterns of variation in the presence or
absence of eQTL, as well as the heterogeneity of effect sizes across tissues.}

{
}

{\color{black} The contribution of the paper is five-fold:
1) introduction of a novel hierarchical Bayesian model for multi-tissue eQTL analysis;
2) development of an efficient EM algorithm for estimating the parameters of the model;
3) analysis of the properties of the model;
4) rigorous theoretical arguments showing that model-based testing procedures control FDR under realistic assumptions;
5) applications to the GTEx data.
}

{

\begin{figure}[htbp]
  \centering
  \begin{tabular}[c]{cc}
    \begin{tabular}[b]{c}
      \subfloat[Raw data format]{\label{fig:left}%
      \includegraphics[width=1.3in,height=3.6in]{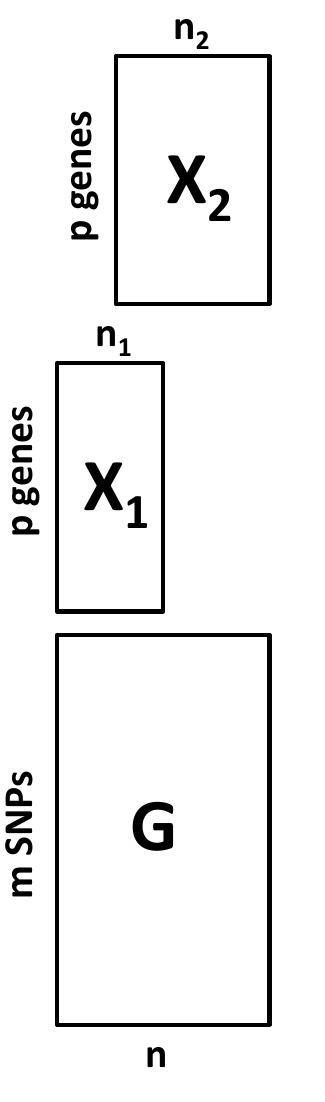}}
    \end{tabular}
    &
    \subfloat[MT-eQTL input and output]{\label{fig:right}%
      \begin{tabular}[b]{c}
        \includegraphics[width=1.85in,height=1.85in]{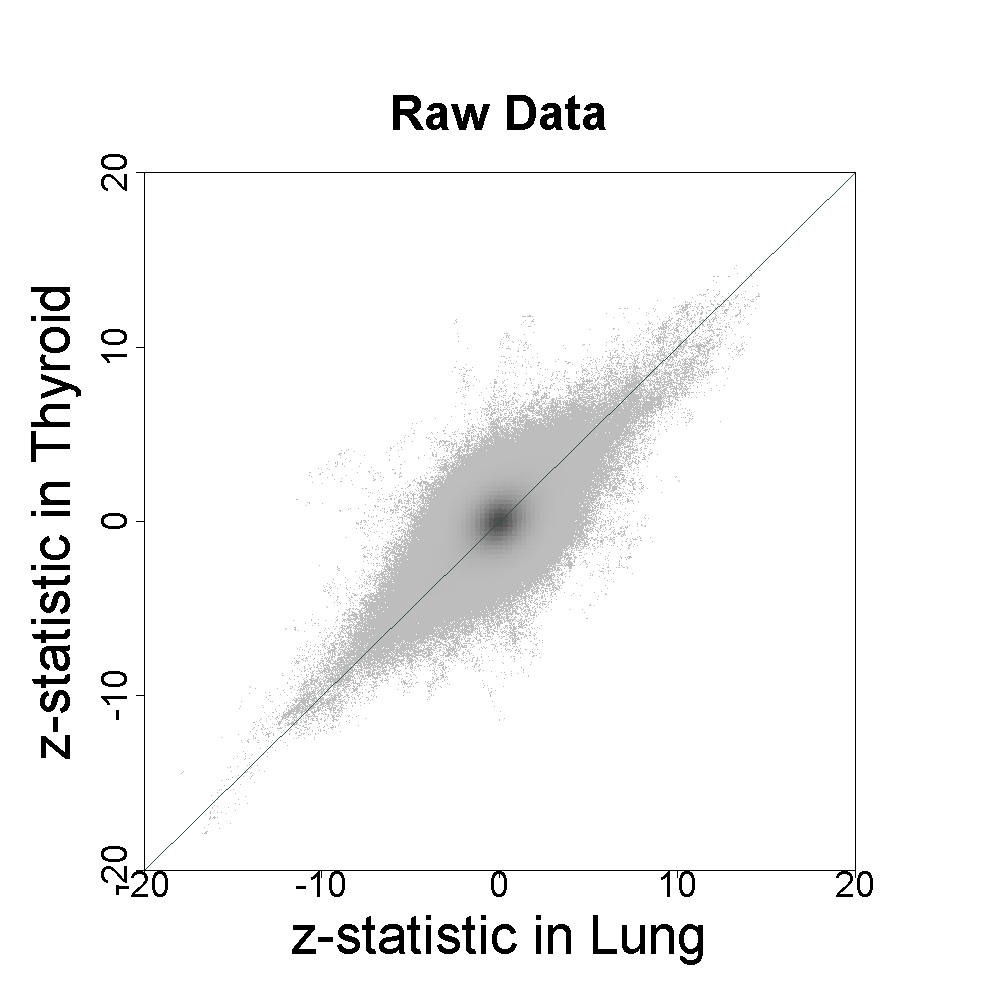}\\[2mm]
        \includegraphics[width=1.8in,height=1.8in]{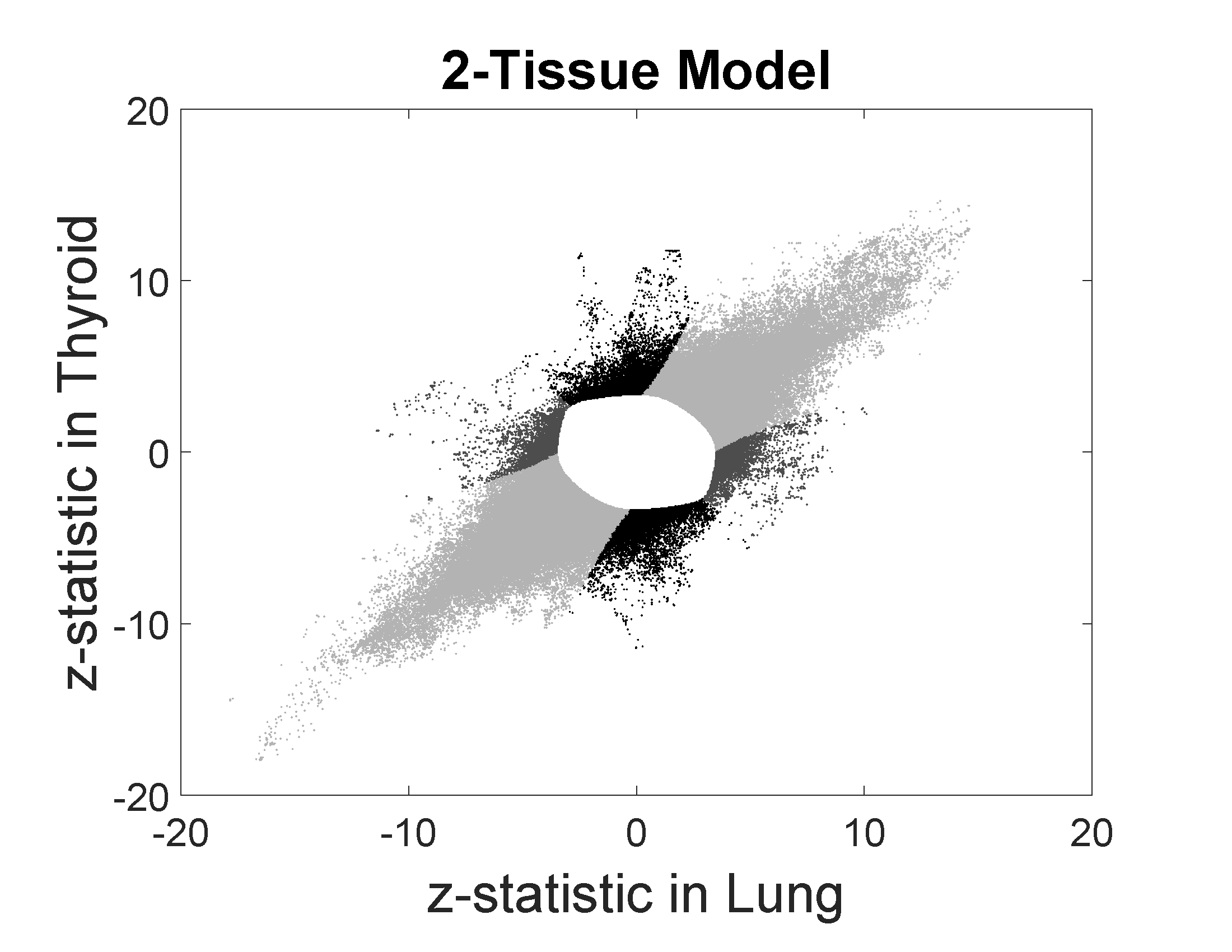}
      \end{tabular}}
  \end{tabular}
  \vskip.6in
  \caption{(a) Illustration of the typical data format with two tissues.
Genotype data $G$ is available for $m$ SNPs and each of $n$ samples.
Expression measurements are available for $p$ genes; sample sets for different tissues may not be the same.
(b) z-statistics for lung and thyroid: density plot for all local gene-SNP pairs (top),
and scatter plot for significant local gene-SNP pairs with tissue specificity by gray scale (bottom). {\color{black}The gene-SNP pairs deemed insignificant are omitted, leading to the white space at the center of the plot.  The remaining points are colored according to their assessed tissue specificity:
dark gray points correspond to the Lung-specific eQTL; black points correspond to the Thyroid-specific eQTL; light gray points correspond to the cross-tissue eQTL.}}
\vspace*{-3pt}
\end{figure}

\subsection{Related work}

Most existing multi-tissue analyses extract eQTL individually from each tissue and then apply post-hoc procedures
to assess commonality and specificity \citep{dimas2009common,fu2012unraveling,nica2011architecture,brown2013integrative}.
Recently, several joint analysis approaches were proposed. 
\citet{gerrits2009expression} used an ANOVA model to study the genotype effect on a transcript across several cell types.
\citet{petretto2010new} used a sparse Bayesian multivariate regression model to identify eQTL at
multiple loci for same transcripts in many tissues.
More recently, \citet{flutre2013statistical} developed a Bayesian model and a permutation-based approach to identify eQTL in multiple tissues. The computation is prohibitive for a moderate number of tissues and a large number of gene-SNP pairs.
\citet{sul2013effectively} proposed a ``Meta-Tissue" method that combines linear mixed models with meta-analysis.
It focuses on one gene-SNP pair at a time.
{\color{black}However, the method cannot borrow strength across gene-SNP pairs for eQTL detection, or provide global parameter estimates to characterize eQTL patterns.}

In the literature, eQTL analyses are generally divided into two categories: gene-level analysis and SNP-level analysis. The former focuses on the identification of eQTL genes, typically by averaging evidence over all candidate SNPs. The latter treats all gene-SNP pairs equally and aims at identifying significantly associated pairs.
Both \citet{gerrits2009expression} and \citet{sul2013effectively} studied eQTL at the SNP level while \citet{petretto2010new} and \citet{flutre2013statistical} are gene-level studies.
Gene-level analysis tries to address linkage disequilibrium by assuming there is at most one causal SNP for each gene.
However,
it cannot provide a list of candidate SNP loci which are potential eQTL for a gene.
In this paper, we shall focus on the SNP level study, providing a complementary view of the problem.
We will also address the computational issue and the lack-of-power concern by exploiting an empirical Bayes approach. 

%

\section{The MT-eQTL Model}\label{sec:model}


\subsection{Format of Multi-Tissue eQTL Data}

The general data format for the multi-tissue eQTL problem is as follows.
For each of $n$ donors we have full genotype information, and measurements of gene expression
in at least one of $K$ tissues.
Let ${\bf G}$ be an $m \times n$ matrix containing the measured genotype of each donor in the study
at $m$ SNPs.
The entries take values $0$, $1$, and $2$, typically coded as the number of minor allele variants.
Each column of ${\bf G}$ corresponds to a donor, and each row
corresponds to a SNP.
The measured transcript levels for tissue $k$ are contained
in a $p \times n_k$ matrix ${\bf X}_k$, where $p$ is the number of genes, and $n_k \leq n$
is the number of donors for tissue $k$.
The number of donors $n_k$ can vary widely among tissues,
and even if two tissues have similar numbers of samples, they may have  few common
donors.  The data available for the purposes of multi-tissue eQTL analysis has the form
$({\bf G}, {\bf X}_1, \ldots, {\bf X}_K)$.
Figure \ref{fig:left} gives an illustration of the typical data format with two tissues.



In most cases eQTL analysis is preceded by several preprocessing steps and covariate adjustment.
Covariate adjustment is necessary because genotype and expression data usually contain confounding factors.
Some confounders, such as gender, are observed, while others are of unknown technical or biological origin.
To identify the unknown confounding factors, most studies use principal components, surrogate variables \citep{leek2007capturing},
or PEER factors \citep{stegle2012using} as covariates.
In Section \ref{GTExproc}, we shall discuss the preprocessing procedure of the GTEx data.
For now, we just assume that the expression data and genotype data have been appropriately residualized for confounders,
so the comparison of these residualized quantities are partial correlations adjusted for covariates.

\subsection{Multivariate z-Statistic from Single Tissue Correlations}\label{sec:notation}

Denote a gene by $i \in \{ 1, \ldots, p \}$ and a
SNP by $j \in \{1, \ldots, m \}$.
We focus on a subset $\Lambda$ of the
full index set $\{ 1, \ldots, p \} \times \{1, \ldots, m \}$ that consists of pairs $(i,j)$ such that SNP
$j$ is located within a fixed distance (usually 100 Kilobases or 1 Megabase)
of the transcription start site of gene $i$.

Let $\lambda = (i,j)$ be a gene-SNP pair of interest.
Let $r_{\lambda k}$ and $\rho_{\lambda k}$ denote, respectively, the sample and population correlation
of transcript $i$ and SNP $j$ in tissue $k$.
{We use the Pearson product-moment correlation for several reasons: 1) with proper transformation of transcript data, the sample correlation has a known, normal distribution \citep{winterbottom1979note}, which is the basis of the proposed multi-tissue model;
2) the Pearson correlation has close connection with the regression coefficient in a simple linear
model relating transcript abundance and genotype (the foundation of most single-tissue eQTL studies).}
Note that the sample correlation $r_{\lambda k}$ depends only on the $n_k$ measurements
from donors of tissue $k$.
The vector of correlations $\br_{\lambda} = (r_{\lambda 1}, \ldots, r_{\lambda K})$ captures the association between the expression of transcript $i$ and the value of genotype $j$ in $K$
tissues.  Relationships between different tissues
will be reflected in correlations between the entries of $\br_{\lambda}$.
These features make
$\br_{\lambda}$ a natural starting point for a multi-tissue eQTL model.

We build a multivariate model for the correlation vector $\br_{\lambda}$.
Let
$
\bh(\br_{\lambda}) \ = \ \big( h(r_{\lambda 1}), \ldots, h(r_{\lambda K}) \big)
$
be the vector obtained by applying the Fisher transformation
$
h(r) = \frac{1}{2} \log \Big(\frac{1 + r}{1-r}\Big) 
$
to each component of $\br_{\lambda}$.  Let
$
{\bf d}^{1/2} := ( \sqrt{d_1 - 3}, \ldots, \sqrt{d_K - 3} )
$
be a scaling vector, where $d_k$ is the degrees of freedom for $\bX_k$ and $\bG$,
equal to $n_k$ minus the number of covariates used to correct
genotype and expression for samples in tissue $k$.
Finally, define the vector
$
\bz_{\lambda} \ = \ {\bf d}^{1/2} \cdot \bh(\br_{\lambda})
$
where $\bu \cdot \bv$ denotes the Hadamard (entry-wise) product of vectors $\bu$ and $\bv$.
{\color{black}Let $\bZ_\lambda$ denote the random vector for $\bz_{\lambda}$.}
If we assume that the expression measurements ${\bf X}_k$ are approximately normal, standard arguments for the Fisher transformation \citep{winterbottom1979note} imply that $h(r_{\lambda k})$ is approximately
normal with mean $h(\rho_{\lambda k})$ and variance $(d_k - 3)^{-1}$.
By a routine multivariate extension of this fact, $\bZ_{\lambda}$ is approximately normally
distributed with mean
$
\bmu_{\lambda} \ = \ {\bf d}^{-1/2} \cdot \bh(\brho_{\lambda}).
$
The variance stabilizing property of the Fisher transformation and our choice of scaling ensures that the
variance of each entry $Z_{\lambda k}$ of $\bZ_{\lambda}$ is close to one, regardless of $\brho_{\lambda}$.
In particular, if the true correlation $\rho_{\lambda k}$ between transcript $i$ and SNP $j$ for tissue $k$
is zero, then $Z_{\lambda k}$ is approximately standard normal.   Thus the $k$-th
entry of the observed vector $\bz_{\lambda}$ is a z-statistic for testing $\rho_{\lambda k} = 0$
vs.\ $\rho_{\lambda k} \neq 0$.

{\color{black} The use of z-statistics greatly reduces the data complexity and magnitude, without losing  much information regarding gene-SNP associations. It facilitates statistical modeling and computation.}  
Importantly, the components of $\bZ_{\lambda}$ are not
independent due to the correlation of effect sizes and  sample overlaps in different tissues.
Capturing this dependence is one of the key
features of the MT-eQTL model, which is described in detail below.

\subsection{Hierarchical Model}

Let $\lambda = (i,j)$ be a gene-SNP pair in $\Lambda$.
MT-eQTL is a multivariate, hierarchical Bayesian model for
the random vector $\bZ_{\lambda}$.   In detail, we assume that
\bea
\label{model1}
\bZ_{\lambda} \, | \, \bmu_{\lambda} & \sim & {\cal N}_K \left( \bmu_{\lambda}, \Delta \right), \\
\label{model2}
\bmu_{\lambda}  & = & \bGam_{\lambda} \cdot \balpha_{\lambda}, \\ 
\label{model3}
\bGam_{\lambda}  & \sim & \bp \mbox{ on } \{0,1\}^K, \\
\label{model4}
\balpha_{\lambda} & \sim & {\cal N}_K( \bmu_0, \Sigma),  \mbox{ independent of } \bGam_{\lambda}.
\eea

{
We briefly explain the rationale behind the model setup.
The first relation is a consequence of the Fisher transformation, where
$\bmu_{\lambda}$ denotes the true effect sizes of the gene-SNP pair $\lambda$ across the $K$ tissues.
The $K \times K$ covariance matrix $\Delta$ has diagonal values 1; its off-diagonal values capture the correlations
between any two tissues arising from the underlying sampling process.
In practice, the off-diagonal values are typically weakly positive due to overlapping donors for different tissues.
Since the true effect sizes are unknown in practice, in \eqref{model2}, we build a hierarchical Bayesian model for  $\bmu_{\lambda}$ based on two assumptions: {\color{black}when the SNP has no effect on the gene in a tissue, the true effect size is 0; when the SNP regulates the gene in a tissue, the true effect size follows a random distribution.}
Thus $\bmu_{\lambda}$ is represented as a Hadamard product of two random vectors,
$\bGam_{\lambda}$ and $\balpha_{\lambda}$.
}

{
The random vector $\bGam_{\lambda}$ is a configuration vector for the gene-SNP pair $\lambda$, indicating whether there is an eQTL in each of the K tissues.
As in \eqref{model3}, the prior distribution of $\bGam_\lambda$  is a multinomial distribution with $\bp$ being the probability mass function.
The multinomial distribution has $2^K$ components, each being a length-$K$ vector of $0's$ and $1's$.
In particular, $\bGam_\lambda=\0$ indicates there is no eQTL in any tissue for the gene-SNP pair $\lambda$, and $\bGam_\lambda=\mathbf{1}$ indicates there are eQTL in all tissues for this particular gene-SNP pair.
The random vector $\balpha_{\lambda}$ is an eQTL effect size vector for the gene-SNP pair $\lambda$, capturing the true effect size in each tissue if there is an eQTL.
In \eqref{model4}, we give $\balpha_{\lambda}$ a Gaussian prior, with mean $\bmu_0$ and covariance $\Sigma$.
The mean parameter $\bmu_0$ is a length-$K$ vector capturing the average eQTL effect sizes in all tissues, and the $K\times K$ matrix $\Sigma$ represents the covariance structure of eQTL effect sizes across multiple tissues.
The diagonal values indicate the variation of effect sizes in different tissues; and the off-diagonal values, typically strongly positive, reflect the relations of effect sizes between tissues.
}



{
In the model, there are four major parameters, $\Delta$, $\bp$, $\bmu_0$ and $\Sigma$.
The parameters characterize multi-tissue effect sizes for all gene-SNP pairs, and carry important biological interpretations.
We will exploit an empirical Bayes approach to estimate all parameters from data.
}


\subsection{Mixture Model and Estimation}\label{est}

The hierarchical model (\ref{model1})-(\ref{model4}) describing the distribution of $\bZ_\lambda$ is fully specified by
$\theta = (\bmu_0, \Delta, \Sigma, \bp)$, which consists of $2^K + K^2 + K -1$ real-valued parameters.
Estimation of, and inference from, the hierarchical model is based on an equivalent mixture representation. 

If ${\bf U}$ is distributed as ${\cal N}_K(\bmu,\Sigma)$ and $\bgam$ is a fixed vector in $\{0,1\}^K$,
then one may readily verify that the entrywise product ${\bf U} \cdot \bgam$ is distributed as
${\cal N}_K \big( \bmu \cdot \bgam, \Sigma \cdot \bgam \bgam^T \big)$.
A straightforward argument then shows that the hierarchical model
(\ref{model1})-(\ref{model4}) is equivalent to  a mixture model 
\begin{eqnarray}
\label{mixture}
\bZ_{\lambda}
\ \sim \
\sum_{\bgam \in \{0,1\}^K} p_{\bgam} \, {\cal N}_K \big( \bmu_0 \cdot \bgam, \, \Delta +
  \Sigma \cdot \bgam \bgam^T \big).
\end{eqnarray}
The mixture model is readily interpretable.
Each component of the model corresponds to a unique configuration $\bgam$, or equivalently,
a unique pattern of tissue specificity.   The model component
corresponding to $\bgam = {\bf 0}$ represents the case in which there are no eQTL in any tissue, and has
associated (null) distribution ${\cal N}_K( {\bf 0}, \Delta)$.  The model component
corresponding to $\bgam = {\bf 1}$ represents the case in which there are eQTL in every tissue, and has
associated distribution ${\cal N}_K( \bmu_0, \Delta + \Sigma)$.  Other values of $\bgam$ represent intermediate cases
in which there are eQTL in some tissues (those with $\gamma_k = 1$) and not in others (those with $\gamma_k = 0$).

{
We adopt an empirical Bayes approach, estimating the model parameters $\theta = (\bmu_0, \Delta, \Sigma, \bp)$
from the observed z-statistics $\{ \bz_{\lambda} :\lambda \in \Lambda \}$ by maximizing the likelihood derived from (\ref{mixture}).
Beginning with the work of \citet{newton2001differential} and \citet{efron2001empirical},
empirical Bayes approaches have been applied to hierarchical models in a number of genetic applications,
most notably the study of differential expression and co-expression in gene expression microarrays
\citep{newton2004detecting,smyth2004linear,efron2008microarrays,dawson2012empirical}.
}

{\color{black} Directly maximizing the joint log likelihood of the model \eqref{mixture} across gene-SNP pairs is computationally intractable.
On the one hand, observations for different gene-SNP pairs may be correlated, as each gene may contain multiple SNPs and neighboring SNPs may have relatively strong linkage disequilibrium.
On the other hand, the likelihood function for each gene-SNP pair has $2^K$ components, each corresponding to a weighted multivariate Gaussian likelihood function with overlapping model parameters.
Note that the parameters in the model \eqref{mixture} determine, and are determined by,
the {\em marginal} distribution of the vectors $\bZ_\lambda$, and do not depend on their joint distribution.
We address the issue of correlated observations by maximizing a marginal composite likelihood, which is defined as the product of the marginal likelihoods of all considered gene-SNP pairs.  As such,
it does not attempt to capture correlation between different gene-SNP pairs.
For typical eQTL analyses, in which the number of gene-SNP pairs is large and average pairwise correlations are low, we expect the use of marginal composite likelihood estimation has little effect on statistical efficiency.
}

{\color{black}
To address the difficulty of parameter estimation, we exploit an EM algorithm by treating the underlying configuration vector for each gene-SNP pair as a latent variable.
As a result, the estimation of the probability mass function $\bp$ can be separated from the estimation of $\bmu_0$, $\Delta$ and $\Sigma$.
The optimization with respect to $\bp$ has a closed-form solution in each iteration.
Furthermore, in cis-eQTL analysis, the null configuration $\bgam=\0$ and the full alternative configuration $\bgam=\mathbf{1}$ together usually account for the majority of the prior weight.
When estimating $\bmu_0$, $\Delta$ and $\Sigma$, if we only focus on the log likelihood terms corresponding to these two configurations, each parameter has an explicit estimate.
As such, we use a modified EM algorithm with the two-term approximation, which greatly reduces the computational cost.
Simulation studies show that such approximation has little affect on the accuracy of the estimation.
More details of the model fitting algorithm can be found in Section 1 of the online supplementary material. }

\subsection{Marginal Compatibility}


In eQTL studies with multiple tissues, it is desirable if the model for a subset of tissues is compatible with the model for a superset of tissues in the sense that the former can be obtained from the latter via marginalization.
We refer to this property as {\it marginal compatibility}.
From the model interpretation point of view, the property guarantees that parameters (e.g., prior probabilities of different eQTL configurations, covariance of effect sizes in different tissues) corresponding to a set of tissues do not depend on whether we observe just those tissues or a superset of the tissues.
It is crucial in multi-tissue eQTL studies as we essentially always analyze a set of some hypothetical superset of tissues that we do not observe.
From the model fitting point of view, with the property, we only need to fit the full model with all available tissues once. The model for any subset of tissues can be obtained directly through marginalization without refitting.


To elaborate, let $S \subseteq \{1,\ldots, K\}$ be a subset of $r$ tissues, with $1 \leq r \leq K$.
The mixture model (\ref{mixture}) has two important compatibility properties:
(i) the marginalization of the full model to $S$ has the same general form as the model derived
from $S$ alone; and (ii) the parameters of the marginal model are obtained by restricting the
parameters of the full model to $S$.  The following definition and lemma makes these statements
precise.  See Section 2 in the online supplementary material for a proof of the lemma.

\noindent
{\bf Definition:} Let $S \subseteq \{1,\ldots, K\}$ with cardinality $|S| = r$.  For each vector ${\bf u} \in \real^K$ let
${\bf u}_{\scriptscriptstyle S} = (u_k : k \in S) \in \real^r$ be the vector obtained by restricting ${\bf u}$ to the
entries in $S$.  Similarly, for each matrix $A \in \real^{K \times K}$ let
$A_{\scriptscriptstyle S} = \{ a_{kl} : k,l \in S \}$ be the $r \times r$ matrix obtained by retaining only the rows and columns
with indices in $S$.  Note that if $A$ is non-negative (positive) definite, then $A_{\scriptscriptstyle S}$
is non-negative (positive) definite as well.

\begin{lem}
\label{compat}
If $\bZ \in \real^K$ be a random vector having the mixture distribution (\ref{mixture}), then
\bes
\label{mixmarg}
\bZ_{\scriptscriptstyle S}
\ \sim \
\sum_{\bzeta \in \{0,1\}^r} p_{\scriptscriptstyle S, \bzeta} \,
{\cal N}_r \big( \bmu_{0 \scriptscriptstyle S} \cdot \bzeta, \,
\Delta_{\scriptscriptstyle S} + \Sigma_{\scriptscriptstyle S} \cdot \bzeta \bzeta^T \big)
\ees
where $(p_{\scriptscriptstyle S, \0},\cdots,p_{\scriptscriptstyle S, \bf 1})$ is the probability mass function on  $\{0,1\}^r$
obtained by marginalizing $\bp$ to $S$, i.e., $p_{\scriptscriptstyle S, \bzeta}=\sum_{\bgam:\bgam_S=\bzeta}p_{\bgam}$.
\end{lem}


\section{Multi-Tissue eQTL Inference}
\label{MTI}

Once fit, the mixture model (\ref{mixture}) provides the basis for inference about eQTL across tissues.
{\color{black} When the number of gene-SNP pairs is large, as in the GTEx example in Section \ref{GTEx}, $\theta$ can be accurately estimated from data.}
At the level of posterior inference for gene-SNP pairs, we therefore regard $\theta$ as fixed and known.
For data sets with small sample sizes, approximate standard errors for the components of $\theta$ can be obtained from the likelihood via the observed information matrix.

Denote the density of the distribution
${\cal N}_K \big( \bmu_0 \cdot \bgam, \, \Delta + \Sigma \cdot \bgam \bgam^T \big)$
associated with the configuration $\bgam \in \{0,1\}^K$
by $f_{\bgam}$.
Thus under the mixture model (\ref{mixture}) the random vector $\bZ_{\lambda}$ has density
$f(\bz) \ = \ \sum_{\bgam} p_{\bgam} \, f_{\bgam}(\bz)$, $\bz \in \real^K$.
In view of this expression and the hierarchical model (\ref{model1})-(\ref{model4}), one may
regard $\bZ_{\lambda}$ as one element of a
jointly distributed pair $(\bGam_{\lambda}, \bZ_{\lambda})$, where
\be
\label{eqn:gamz}
\bGam_{\lambda} \sim \bp
\ \mbox{ and } \
\bZ_{\lambda} \, | \, \bGam_{\lambda}  \, \sim \, f_{\bgam} .
\ee
We carry out multi-tissue eQTL analysis based on the posterior distribution of the
configuration $\bGam_{\lambda}$ given the observed vector of z-statistics $\bz_\lambda$.
Two inference problems are of central interest: one is eQTL detection, in all tissues and in a subset of tissues;
the other is the assessment of eQTL tissue specificity given eQTL is present in at least one tissue.

\subsection{Detection of eQTL Using the Local False Discovery Rate}
\label{sec:lfdr}

A primary goal of multi-tissue analysis is testing each transcript-SNP pair
for the presence of an eQTL in at least one tissue.  This can be formulated
as a multiple testing problem:
\be
\label{mult-test}
\mbox{H}_{0, \lambda}: \bGam_{\lambda} = {\bf 0}
\ \mbox{ versus } \
\mbox{H}_{1, \lambda}: \bGam_{\lambda} \neq {\bf 0}
\ \ \mbox{for} \ \ \lambda \in \Lambda .
\ee
For $\lambda = (i,j) \in \Lambda$ the null hypothesis $H_{0,\lambda}$ asserts that SNP $j$ is not an eQTL for transcript $i$ in any tissue,
while the alternative $H_{1,\lambda}$ asserts that there is an eQTL between $i$ and $j$ in at least one tissue.

The null hypotheses can also be expressed in the form
$
\mbox{H}_{0, \lambda}:  \bZ_{\lambda} \sim {\cal N}_K \big( {\bf 0}, \, \Delta \big) .
$
One may derive a p-value for each $\lambda$ directly from the null distribution, and convert it to
control the overall false discovery rate (FDR) \citep[cf.][]{benjamini1995controlling,storey2003statistical}.
However, this procedure ignores relevant information about
the distribution of $\bZ_{\lambda}$ under the alternative that is contained in the
mixture model.

We address the multiple testing problem (\ref{mult-test}) using the local false
discovery rate introduced by \citet{efron2001empirical} in the context
of an empirical Bayes analysis of differential expression in microarrays.
Other applications of the local false discovery rate to genomic problems can be found in
\citet{newton2004detecting}, \citet{efron2007size}, and \citet{efron2008microarrays}.
To simplify notation, let $(\bGam, \bZ)$ denote a generic pair distributed
as $(\bGam_{\lambda}, \bZ_{\lambda})$.

\noindent
{\bf Definition:}
The {\em local false discovery rate} of an observed z-statistic vector $\bz$ under the model
(\ref{mixture}) is defined by
\be
\label{lfdr}
\eta(\bz)
\ : = \
\pr( \bGam = {\bf 0} \, | \, \bZ = \bz)
\ = \
\frac{ p_{\bf 0} f_{\bf 0} (\bz) }{ f(\bz) } .
\ee

Let $\alpha \in (0,1)$ be a target false discovery rate (FDR) for the multiple testing problem (\ref{mult-test}).
Vectors $\bz$ for which the local false discovery rate $\eta(\bz)$ is small provide
evidence for the alternative $\bGam \neq \0$.
We carry out testing of gene-SNP pairs using a step-up procedure applied to the running average of the ordered local false discover rates \citep{newton2004detecting,cai2009simultaneous}.

\noindent
{\bf Local FDR Step-Up Procedure: Target FDR $= \alpha$}
\begin{enumerate}

\item
Given: Observed $z$-statistic vectors $\{ \bz_\lambda : \lambda \in \Lambda \}$.

\item
Enumerate the elements of $\Lambda$ as $\lambda_1, \ldots, \lambda_N$ so that
$\eta(\bz_{\lambda_1}) \, \leq \, \cdots \, \leq \, \eta(\bz_{\lambda_{N}})$.

\item
Reject hypotheses $\mbox{H}_{0,\lambda_1}, \ldots, \mbox{H}_{0,\lambda_L}$,
where $L$ is the largest integer such that \\$L^{-1} \sum_{l=1}^L \eta(\bz_{\lambda_l}) \leq \alpha$.

\end{enumerate}


\subsection{Theoretical Justification of the Local FDR Approach}

{
In order to better understand the local FDR step-up procedure, and to assess its performance,
it is useful to express the procedure in an equivalent form.
As noted by \citet{efron2001empirical}, the false discovery rate associated with a rejection
region $R \subseteq \real^k$ for the multiple testing problem (\ref{mult-test}) is given by
$\pr(\bGam = {\bf 0} \, | \, \bZ \in R)$.
They establish the following elementary fact, which exhibits a connection
between FDR and local FDR.
}

\begin{prop}
\label{l/fdr}
If $R \subseteq \real^k$ is such that $\pr(\bZ \in R) > 0$,
then $\pr(\bGam = {\bf 0} \, | \, \bZ \in R) = \E ( \eta(\bZ) \, | \, \bZ \in R)$.
\end{prop}

{
As noted above, vectors $\bz$ for which $\eta(\bz)$ is small provide evidence against
$\bGam = \0$, so it is natural to reject $\mbox{H}_{0,\lambda}$ when
$\eta(\bz_{\lambda})$ falls below an appropriate threshold.
Consider rejection regions of the form $R(t) = \{ \bz : \eta(\bz) \leq t \}$ for $t \in (0,1)$.
Given a target false discovery rate $\alpha$, we wish to find $t$ such that
$\alpha = \pr( \bGam = {\bf 0} \, | \, \bZ \in R(t) )$.  By Proposition \ref{l/fdr} this
is equivalent to finding $t \in (0,1)$ such that $F(t) = \alpha$, where
\bes
\label{eqn:Fdef}
F(t)
\ := \
\E ( \eta(\bZ) \, | \, \eta(\bZ) \leq t)
\ = \
\frac{\E[ \eta(\bZ) \, \In(\eta(\bZ) \leq t) ]}{\pr(\eta(\bZ) \leq t)} .
\ees
The empirical analog of $F(t)$ is the ratio
\[
\hat{F} (t)
\ = \
\frac{ \sum_{\lambda \in \Lambda} \eta(\bz_\lambda) \, \In(\eta(\bz_\lambda) \leq t) }
       { \sum_{\lambda \in \Lambda} \In(\eta(\bz_\lambda) \leq t) } ,
\]
which depends only on $\eta(\cdot)$ and the observed vectors $\{\bz_\lambda\}$.
The function $F(t)$ is strictly increasing and continuous (see Section 3.1 in the online supplementary material for proof).
Thus if $F(t)$ and $\hat{F} (t)$ were equal, the local FDR step-up procedure
and the idealized threshold procedure would coincide.
In general, $F(t)$ and $\hat{F} (t)$ will be different, but multiplying the numerator and
denominator of $\hat{F} (t)$ by $|\Lambda|^{-1}$ it is evident that the two functions will
be close if $|\Lambda|$ is large and the dependence among the observed $z$-vectors
is not extreme.  Asymptotic control of the FDR by the step-up
procedure is established in Theorem \ref{thm:FDR} below.
The proof can be found in Section 3 of the online supplementary material.
}

{
Let $\Lambda^* \subseteq {\mathbb N} \times {\mathbb N}$ be an infinite index set,
and let $\Lambda_1, \Lambda_2, \ldots \subseteq \Lambda^*$ be a sequence of
finite subsets of $\Lambda^*$.  Let $\alpha \in (0,1)$ be a target FDR that is
less than the maximum value of $\eta(\bz)$.
For each $n \geq 1$ let
$\{ (\bGam_\lambda, \bZ_\lambda) : \lambda \in \Lambda_n \}$ be jointly
distributed pairs having the same distribution as $(\bGam, \bZ)$.
We wish to assess the performance of the local FDR step-up procedure,
which rejects $H_{0,\lambda}$ when
$\eta(\bZ_\lambda) \leq \hat{\theta}_n =  \sup\{ t : \hat{F}_n(t) \leq \alpha \}$ where
\[
\hat{F}_n (t)
\ = \
\frac{ \sum_{\lambda \in \Lambda_n} \eta(\bZ_\lambda) \, \In(\eta(\bZ_\lambda) \leq t) }
       { \sum_{\lambda \in \Lambda_n} \In(\eta(\bZ_\lambda) \leq t) }
\ \ \ \ 0 < t < 1 .
\]
The number of false discoveries and total discoveries for the procedure are equal to
$
M_n \ = \ \sum_{\lambda \in \Lambda_n} \In(\bGam_\lambda = 0) \,
              \In(\eta(\bZ_\lambda) \leq \hat{\theta}_n)
\ \ \mbox{ and } \ \
N_n \ = \  \sum_{\lambda \in \Lambda_n}
              \In(\eta(\bZ_\lambda) \leq \hat{\theta}_n) .
$
}

\begin{thm}
\label{thm:FDR}
Let $(\bGam, \bZ)$ have joint distribution given by Model (\ref{eqn:gamz}) with
parameters $(\bmu_0, \Delta, \Sigma, {\bf p})$.  Assume that $\Delta$ is positive definite and
that the diagonal entries of $\Sigma$ are positive.
If $\hat{F}_n(t) \to F(t)$ in probability for each $t \in (0,1)$ then $\E M_n / \E N_n \to \alpha$ as $n\rightarrow \infty$.
\end{thm}

{
The ratio of expectations $\E M_n / \E N_n$ is sometimes referred to as the marginal false
discovery rate (m-FDR).
\citet{cai2009simultaneous} established optimality properties and m-FDR control
of several local FDR based testing procedures, including the step-up procedure used here,
under independence and monotonicity assumptions.  However, these assumptions are
typically violated in the setting of interest to us here.
The monotonicity assumption, which in the
present case involves the relationship between the distributions of the local FDR $\eta(\bZ_\lambda)$ under
$H_{0,\lambda}$ and $H_{1,\lambda}$, does not appear to hold.  Moreover, in eQTL data there are typically
significant correlations between nearby SNPs (linkage disequilibrium), leading to
to complex, non-stationary correlations between the gene-SNP based vectors $\bZ_\lambda$.
}

{
Theorem \ref{thm:FDR} makes no explicit assumptions on the joint
distribution of the vectors $\bZ_\lambda$; instead it relies on the relatively weak condition that
$\hat{F}_n(t) \to F(t)$ in probability.  This condition holds, for example, under the (very mild)
assumption that the variance of the numerator and the denominator
of $\hat{F}_n(t)$ are of order $o(|\Lambda_n|^2)$.
{\color{black}The variance decay assumption concerns
the family of all gene-SNP pairs, across all measured genes instead of a single gene.
Although the SNPs co-located with a particular gene may be highly correlated, correlations are generally weak, or zero, across distant genes. These distal pairs dominate the index set $\Lambda_n$, and so the variance decay assumption should be satisfied in any cis eQTL analysis involving a large number of genes.
When the assumption holds, the conclusion of the theorem may
be strengthened to $M_n / N_n = \alpha + o_P(1)$.}
}

\subsection{Analysis for Subsets of Tissues}

In some problems, a subset $S \subseteq \{1,\ldots,K\}$ of the available tissues may be of primary interest.
The multiple testing framework described above can be adapted to the tissues in $S$ in two primary ways.
The first is to construct a model based only on the tissues in $S$ and use the resulting local FDR to identify
multi-tissue eQTL.  However, this approach does not make use of the available data from tissues outside
$S$ and as such it does not borrow strength from commonalities among tissues.  As an alternative,
one may use the {\em marginal local FDR} for $S$, defined by
\be\label{marglfdr}
\eta_{S}(\bz)
\ : = \
\pr( \bGam_S = {\bf 0} \, | \, \bZ = \bz)
\ = \
\frac{ \sum_{\bgam: \bgam_{S} = {\bf 0}} p_{\bgam} f_{\bgam} (\bz) }{ f(\bz) } .
\ee
Here $\bGam_S$ and $\bgam_S$ denote, respectively, the restriction of the vectors $\bGam$ and $\bgam$
to the tissues in $S$, while $p_{\bgam}$, $f_{\bgam}$ and $f$ correspond to the full model (\ref{mixture}).
We emphasize that the marginal local FDR $\eta_{S}(\bz)$ is a function of the complete vector of z-statistics, and therefore depends on the fitted model for the full set of tissues.
{\color{black}In Section \ref{gtexresult}, we have shown that the marginal local FDR derived from the full data set is uniformly more powerful than the local FDR derived from a subset of the data in detecting eQTLs in a subset of tissues. More numerical results can be found in Section 4.3 of the supplementary material.}

\subsection{Assessments of Tissue Specificity}
\label{sec:ts}

Testing gene-SNP pairs is typically the first step in multi-tissue eQTL analysis.
Rejection of $\mbox{H}_{0,\lambda}$
is based on evidence that $\lambda$ is an eQTL in at least one of the available tissues.
More detailed statements about the pattern of eQTL across tissues can be made
using information about the full configuration vector $\bGam_\lambda$.
If the hypothesis $\mbox{H}_{0,\lambda}$ is rejected, a natural estimate of $\bGam_\lambda$
is the maximum a-posteriori (MAP) configuration defined by
\[
\widehat{\bgam}_{\lambda} \ = \ \argmax_{\bgam \in \{0,1\}^K\setminus \0}
p(\bgam \, | \, \bz_{\lambda})
\ = \
 \argmax_{\bgam \in \{0,1\}^K\setminus \0} \ p_{\bgam} \, f_{\bgam} (\bz_{\lambda}).
\]
As an alternative, one may compute the
marginal posterior probability of an eQTL in each tissue $k$, namely
$
p(\bGam_{\lambda,k}=1|\bz_\lambda)\ =\ \sum\limits_{\bgam:\bgam_k=1} p(\bgam|\bz_\lambda)
\ = \
\sum\limits_{\bgam:\bgam_k=1} \ p_{\bgam} \, f_{\bgam} (\bz_{\lambda}) / f(\bz_{\lambda}) ,
$
and declare an eQTL in tissue $k$ if this marginal probability exceeds a predefined threshold.
Both MAP and thresholding of the marginal posterior extend to subsets of tissues.

\subsection{Testing a Family of Configurations}

The goal of the multiple testing problem (\ref{mult-test}) is to determine whether
the configuration $\bGam_\lambda$ of a gene-SNP pair is equal to
${\bf 0}$ or belongs to the complementary set $\{0,1\}^K \setminus \{ {\bf 0} \}$.
More generally, one may test membership of $\bGam_\lambda$ in any fixed
subset $T \subseteq \{0,1\}^K$ of configurations.  The associated testing problem
can be written as
\be
\label{gentest}
\mbox{H}_{0, \lambda}^T : \bGam_{\lambda} \in {T^c}
\ \mbox{ versus } \
\mbox{H}_{1, \lambda}^T : \bGam_{\lambda} \in {T},
\ \ \ \ \lambda \in \Lambda .
\ee
A test statistic for (\ref{gentest}) can be obtained by marginalizing the full local FDR (\ref{lfdr}) as
\[
\eta_T(\bz)\ : = \
\pr( \bGam \in T^c \, | \, \bZ = \bz)
\ = \
\frac{ \sum_{\bgam: \bgam \in T^c} p_{\bgam} f_{\bgam} (\bz) }{ f(\bz) }.
\]
The local FDR step-up procedure can then be applied to the values $\{ \eta_T(\bz_\lambda) \}$ in order to
control the overall FDR in (\ref{gentest}).


\section{GTEx Data Analysis}\label{GTEx}

In this section, we apply the MT-eQTL model and inference procedures to the GTEx pilot data freeze
\citep{ardlie2015genotype}.
A pointer to the publicly available data is at \url{http://www.broadinstitute.org/gtex/}.

\subsection{Data Preprocessing}\label{GTExproc}

We focus on nine primary tissues having between 80 and 160 samples: adipose, artery, blood, heart, lung, muscle,
nerve, skin, and thyroid.   In what follows, tissues will be ordered alphabetically.
In total, there are 175 genotyped individuals with expression data in at least one of these tissues {\color{black}(the sample information can be found in Figure S1 of the supplementary material).}



Each entry of the genotype data matrix $\G$ records the number of minor allele variants of one donor at one SNP locus.
Any missing value at a locus was imputed by the corresponding row average.
Loci with minor allele frequency less than $5\%$ in all genotyped individuals were discarded, resulting in
slightly less than 7 million SNPs.
The expression level for each gene in each tissue and sample is measured by the number of
mapped reads per kilobase per million reads (RPKM).
Genes having fewer than $10$ samples with RPKM greater than $0.1$ in some tissue were discarded, leaving
slightly more than 20 thousand genes.
To improve robustness, the gene expression values across samples in a tissue
were inverse quantile normalized. 

Fifteen PEER factors were identified from the expression data from each tissue, and three principal components
were identified from the genotype data.  With an additional covariate for gender, we obtained nineteen covariates
in total.
For each tissue, the confounding effects were adjusted by residualizing the
expression data and the corresponding genotype data on nineteen covariates respectively.
Consequently, the degree of freedom for each tissue is equal to the sample size in that tissue minus 19.

\subsection{Model Fit}
We focus on testing of cis-eQTL, restricting our attention to SNPs that lie within 100 kilobases of
the transcription start site of a gene, yielding roughly 10 million gene-SNP pairs of interest.
Subsequently, the full 9-dimensional MT-eQTL model was fit using the modified EM algorithm described in Section~\ref{est}
with the parameter $\bmu_0$ set to zero.
{\color{black}Fitting the full model took less than 24 hours, and required less than 8 gigabytes of RAM, on a desktop computer with 2.93GHz Intel Xeon CPU.
A comparison of timing results for fitting sub-models of different dimensions between our method and the Meta-Tissue method \citep{sul2013effectively} can be found in Section 5 of the supplementary material.}



In what follows we denote the estimated model parameters by $\theta = (\Delta, \Sigma, \bp)$.
Values of the estimated parameters can be found in Section 5 of the online supplementary material.
The off-diagonal values of $\Delta$ are all positive but small in scale (between $0.07$ and $0.2$),
suggesting that donor overlap among tissues and other features of the experimental design
have a weak but positive effect on the correlations of effect sizes across tissues.
The diagonal values of $\Sigma$ indicate modest heterogeneity of effect size variation across tissues.
The off-diagonal values of $\Sigma$
reflect positive, often large, correlation of effect sizes arising from commonalities among tissues.

The fitted probability mass function $\bp$ assigns probabilities to each of the $2^9$ possible
eQTL configurations.  The most likely configuration is ${\bf 0}$ with $p_\0=0.6808$,
indicating that about $68\%$ of the gene-SNP pairs do not have an eQTL in any tissue.
{\color{black}This is consistent with previous studies \citep{wright2014heritability}.}
{To summarize $\bp$, we sum up the prior probabilities of configurations with the same Hamming weight (defined as the number of 1s in a length-9 binary configuration sequence). 
{\color{black} This provides an overview of the overall probability of seeing an cis-eQTL in $k$ tissues, where $k$ ranges from $0$ to $9$.}
We note, however, that the probabilities for configurations with the same Hamming weight may be quite different.
The total prior probabilities are shown in Figure \ref{fig:P} in the log scale.
{\color{black}The U-shape curve indicates that for cis-eQTL analysis, the most likely configurations are eQTL in no tissue, in a single tissue, or in all tissues, and the least likely configurations are those with eQTL in roughly half the tissues. We remark that the pattern may only apply to cis-eQTL but not to trans-eQTL.}
}

\begin{figure}[htbp]
\begin{centering}
\includegraphics[width=3.5in]{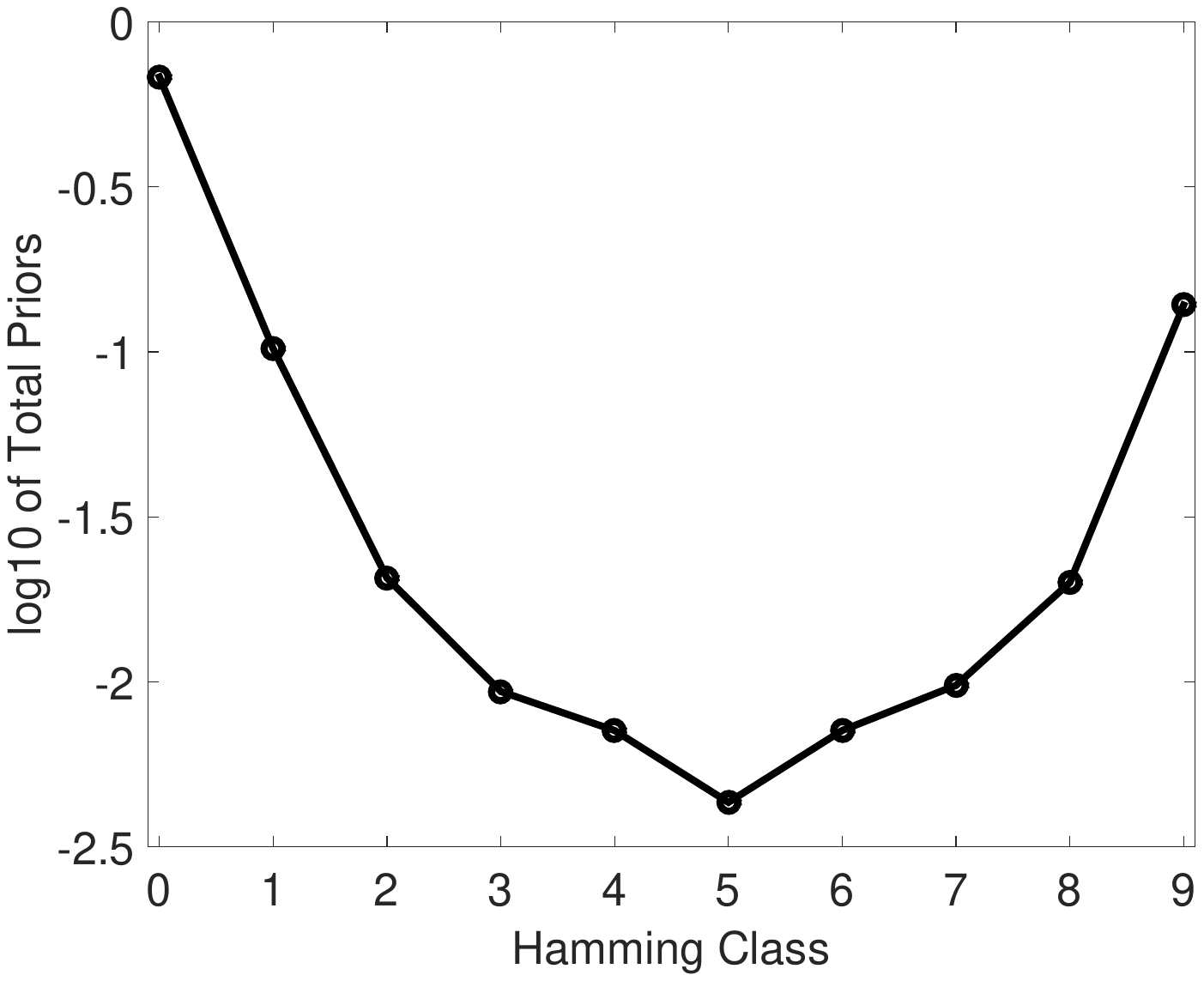}
\caption{Summary of the estimated eQTL probabilities from {\color{black} the cis-eQTL analysis of} the GTEx data. Each circle represents the log (base 10) of the probability of a gene-SNP pair having eQTL in $k$ out of 9 tissues, where $k$ ranges from $0$ to $9$. }
\label{fig:P}
\end{centering}
\end{figure}

\subsection{Results}\label{gtexresult}

Applied to the full 9-dimensional model with FDR threshold $\alpha = .05$, the local FDR step-up procedure
identified roughly 1.28 million gene-SNP pairs (roughly $12\%$ of the total) with an eQTL in at least one
tissue.  We subsequently applied the MAP rule to each significant discovery in order to assess tissue
specificity.
To validate the discoveries, we also applied the Meta-Tissue method to the same data set. Meta-Tissue produces a p value for each gene-SNP pair for testing the existence of eQTL in any tissue.
We further adjusted the p values \citep{benjamini2001control} to control the FDR.
About 80\% of the MT-eQTL discoveries (i.e., 1.03 million) are replicated in Meta-Tissue, providing a highly concordant result.
{\color{black}We further investigated the unique discoveries of each method (about 250 thousand from MT-eQTL, and 177 thousand from Meta-Tissue).}
The left panel of Figure \ref{fig:unique} shows the Meta-Tissue p values of the unique discoveries from MT-eQTL. Small p values are enriched, indicating the unique MT-eQTL discoveries are well supported by Meta-Tissue.
The right panel of Figure \ref{fig:unique} presents the MT-eQTL local FDRs of the unique discoveries from Meta-Tissue.
{\color{black}The unique Meta-Tissue discoveries are only moderately supported by MT-eQTL.
The MT-eQTL provides a systematic way to leverage information across gene-SNP pairs, and offers explicit estimates of model parameters with critical biological interpretation.  }


\begin{figure}[htbp]
\begin{centering}
\includegraphics[width=2.8in]{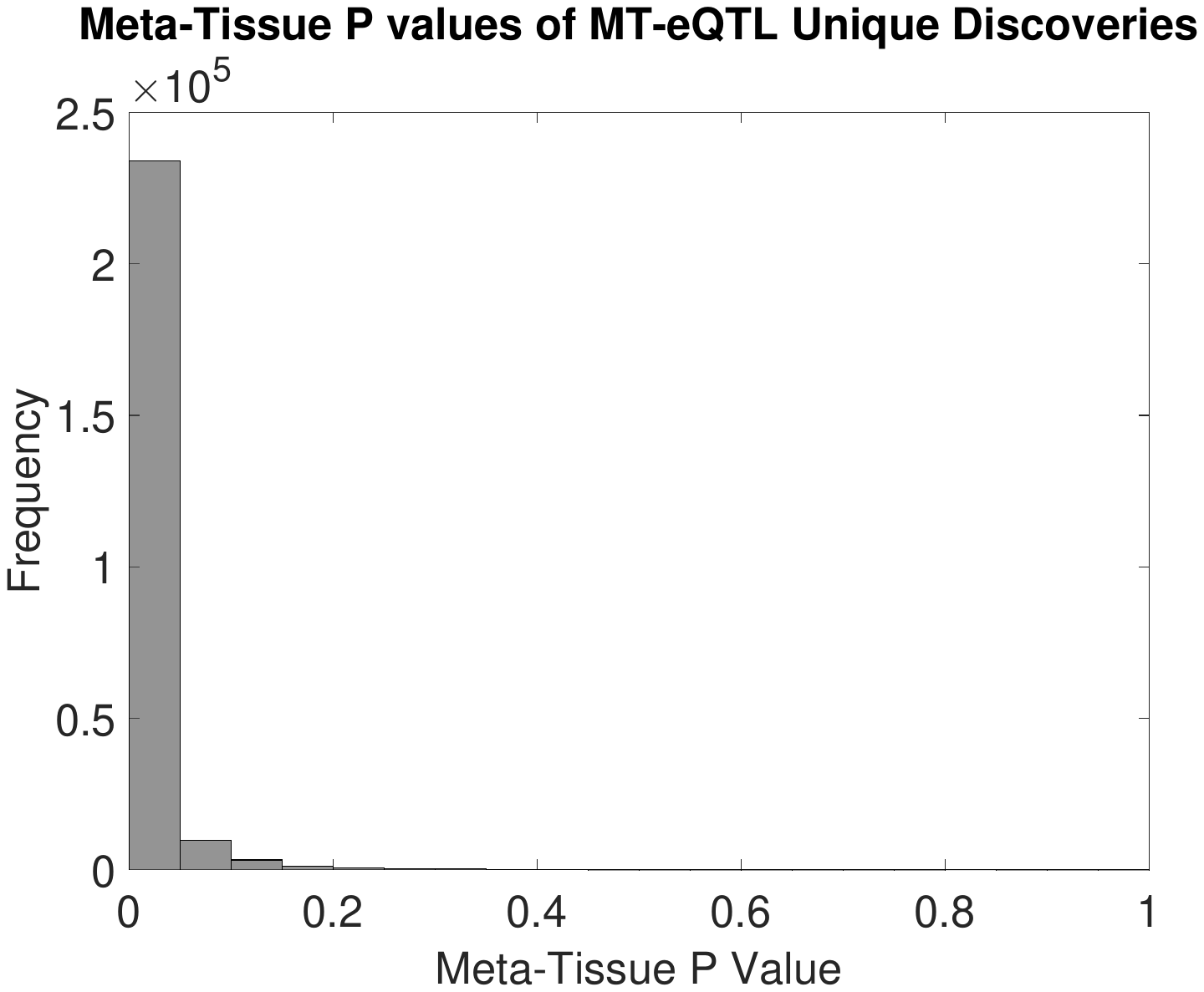}
\includegraphics[width=2.8in]{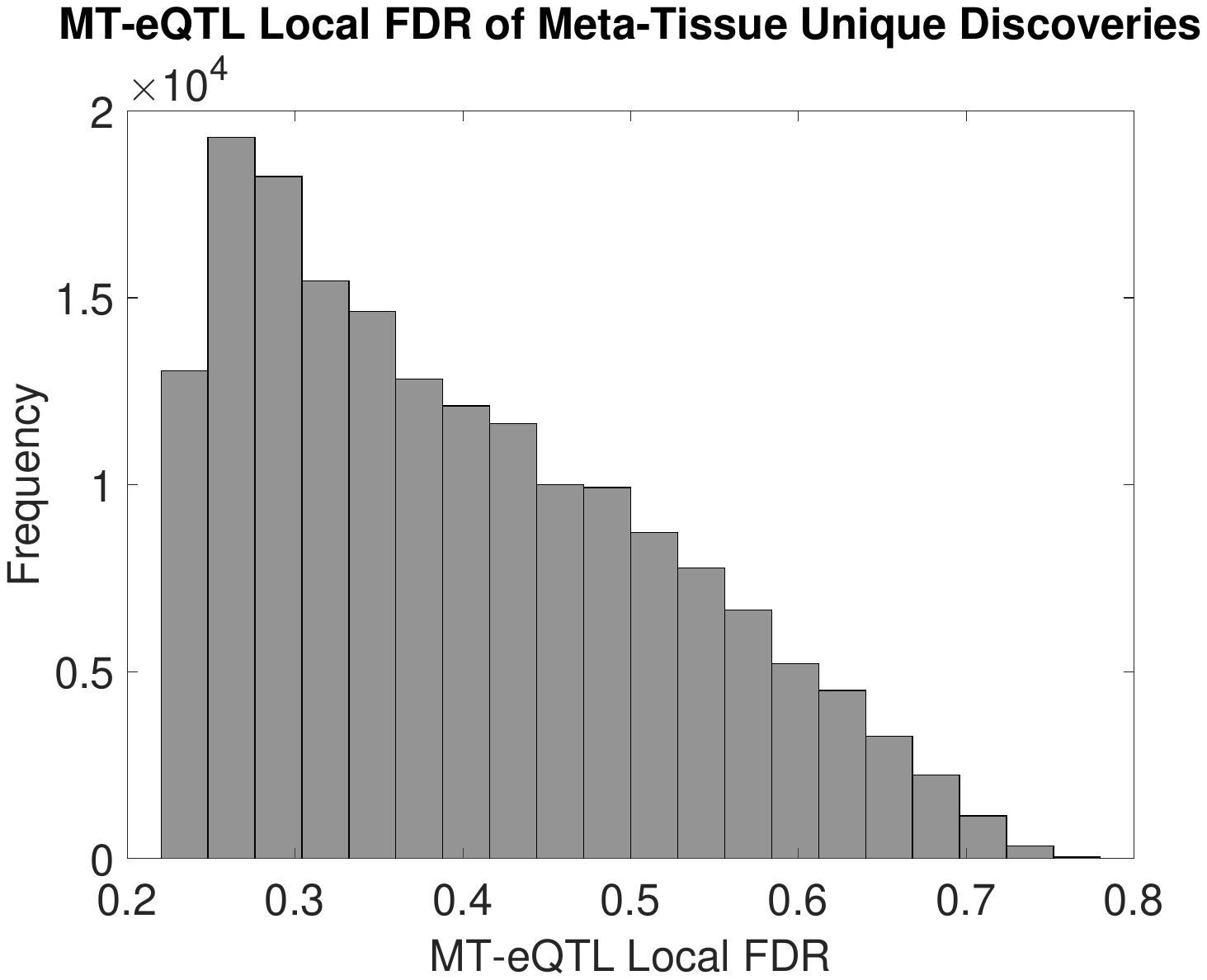}
\caption{The left panel is the histogram of the Meta-Tissue p values for the 250 thousand unique discoveries from MT-eQTL, from the GTEx analysis of eQTL in at least one tissue; The right panel is the histogram of the MT-eQTL local FDRs for the 177 thousand unique discoveries in Meta-Tissue.}
\label{fig:unique}
\end{centering}
\end{figure}

{\color{black}A unique advantage of MT-eQTL over Meta-Tissue is the ease of eQTL tissue specificity assessment.}
To facilitate the visualization of eQTL discoveries, let us focus on a two-tissue MT-eQTL model.
As an example, Figure \ref{fig:right} shows scatter plots of z-statistics for lung and thyroid.
The upper panel shows the density plot of the raw z-statistics (MT-eQTL input);
the lower panel only shows the discoveries with eQTL in at least one of the tissues (MT-eQTL output).
The z-statistic vectors deemed insignificant are omitted, leading to the white space at the center of the plot.  The remaining points are colored according to their assessed tissue specificity based on the MAP approach:
dark gray represents the configuration $(1,0)$ in which there is an eQTL in tissue 1 but not tissue 2;
black represents the configuration $(0,1)$ in which there is an eQTL in tissue 2 but not tissue 1;
and light gray represents the configuration $(1,1)$ in which there is an eQTL in both tissues.
The overall shape of each plot is a tilted ellipse, with extreme values
along the main diagonal and, to a lesser extent, along the coordinate axes.
As expected, significant points close to one of the coordinate axes show
evidence for an eQTL in a single tissue (tissue specific eQTL),
while those along the positive diagonal show evidence for eQTL
in both tissues (common eQTL).  
We remark that this analysis easily extends to an arbitrary number of tissues.

MT-eQTL also effectively leverages information in multiple tissues to improve eQTL detection in a single or a subset of tissues.
To investigate how the use of auxiliary tissues increases statistical power, we studied a sequence of nested MT-eQTL models and focused on eQTL discoveries in a single tissue.
For each of the nine tissues, we first fitted the 1-dimensional model with just the primary tissue and then added other tissues one by one alphabetically to get a sequence of super-models.
For each considered model, we applied the adaptive thresholding procedure to the marginal local FDR for the primary tissue,
and recorded the number of significant discoveries in that tissue.
Figure \ref{fig:line} shows the number of significant discoveries versus the dimension of a model.
Each curve corresponds to a case where one of the nine tissues is set to be the primary tissue.
The number of eQTL discoveries in each primary tissue increases with the dimension of a model.

\begin{figure}[htbp]
\begin{centering}
\includegraphics[width=4in]{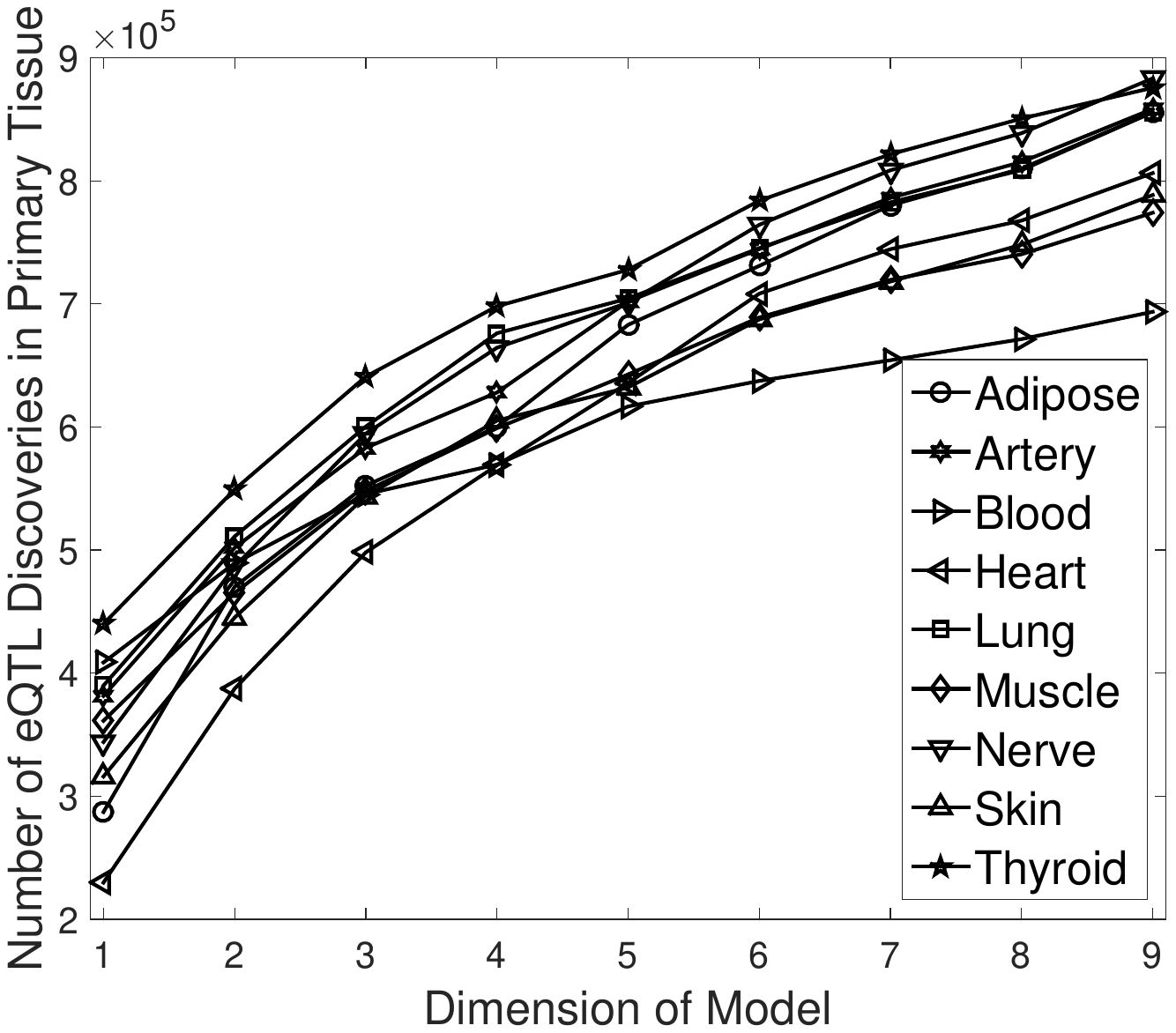}
\caption{The number of significant discoveries in a primary tissue versus the dimension of a MT-eQTL model. Each curve corresponds to a case where one of the nine tissues is set to be the primary tissue. The FDR threshold is fixed to be $0.05$. }
\label{fig:line}
\end{centering}
\end{figure}

\section{Conclusion}\label{sec:conc}

In this paper, we proposed a hierarchical Bayesian model, MT-eQTL, for multi-tissue eQTL analysis.
We adopted an empirical Bayes approach to estimate the model and to perform inferences.
We also proved a substantial theoretical property to support the method in a realistic setting.
The proposed methodology greatly enhances classical single-tissue eQTL analysis methods by accounting for the information shared among tissues.

{\color{black} There are a number of interesting directions for future research. Perhaps the most important is to extend the proposed framework to a large number (e.g., $K\geq 10$) of tissues. The large tissue setting poses real challenges as the total number of configurations grows exponentially in the number of tissues, making the current implementation excessively slow and computationally costly. Another direction is to relax the assumption that the covariance matrix $\Delta$ in Model \eqref{mixture} is constant across gene-SNP pairs. Different genes may have distinct correlation patterns between tissues, which might warrant the use of gene-specific covariance matrices in setting where the number of samples is large. Lastly, it is of interest to extend the method to the identification of trans-eQTLs, which exhibit higher levels of tissue-specificity than cis-eQTLs \citep{jo2016distant}. }

\section*{Acknowledgements}
We would like to thank all the members of the GTEx consortium.
We also thank Dereje Jima for conducting the Meta-Tissue analysis on the GTEx data. 

\section*{Funding}
This work was funded in part by National Institute of
Health (NIH) grants R01 MH090936 and MH101819-0, National Science Foundation (NSF) grants DMS 0907177 and DMS 1310002, and Environmental Protection Agency (EPA) grant STAR RD83580201.

\clearpage

\appendix{Supplementary Materials}

\section{Model Fitting and Parameter Estimation}\label{sec:alg}
\label{MFPE}

\subsection{Matrix eQTL}
\label{sec:matrix eQTL}

The set of correlations $r_{\lambda k}$ for all transcript-SNP pairs $\lambda$ and tissues $k = 1,\ldots, K$ can be conveniently
calculated using the R package Matrix eQTL by \cite{shabalin2012matrix}. The package is designed for fast eQTL analysis in individual tissues. Matrix eQTL accounts for covariates and can filter transcript-SNP pairs by the distance between their genomic  locations. Once Matrix eQTL is applied separately for each tissue, the t-statistics it reports can be transformed into correlations using the simple transformation
\[
	r_{\lambda k}= \frac{t_{\lambda k}}{\sqrt{d_k + t_{\lambda k}^2}}
\]
where $d_k$ is the number of degrees of freedom in the tests for tissue $k$ and is also reported by Matrix eQTL. The set of correlations can then be combined in a single matrix with rows $\br_{\lambda}$.

\subsection{Modified EM Algorithm}
\label{sec:EM}

We wish to estimate the parameter
$\theta = (\bmu_0, \Delta, \Sigma, \bp)$ from the observed z-statistics
$\{ \bz_\lambda : \lambda \in \Lambda \}$, which are computed directly from the sample
correlations $r_{\lambda k}$ obtained from Matrix eQTL.
In order to make the estimation of $\theta$ tractable, we assume that the random vectors $\bZ_\lambda$ are
independent.  The likelihood of the model then has a simple product form,
depending only on the unknown parameter $\theta$, and the observed z-statistics $\{\bz_{\lambda}\}$:
\begin{eqnarray}
\label{likelihood}
L(\{\bz_{\lambda}\}|\theta) =
\prod_{\lambda\in\Lambda} \sum_{\bgam \in \{0,1\}^K} p_{\bgam} \, f_{\bgam}( \bz_{\lambda} \, | \, \theta),
\end{eqnarray}
where $f_{\bgam}(\cdot \, | \, \theta)$ is the probability density function of the
$\mathcal{N}_K \big(\bmu_0  \cdot \bgam,\Delta + \Sigma \cdot \bgam\bgam^T \big)$ distribution.

\noindent
{\bf Remark:}
It is important to note that the parameter $\theta$ concerns only the (common) marginal distribution
of the random vectors $\bZ_\lambda$, and is unaffected by their dependence.  The assumption that the
random vectors $\bZ_\lambda$ are independent facilitates estimation of $\theta$, but does not impose
any constraints on the marginal dependence structure of $\bZ_\lambda$.

We estimate the parameter $\theta$ by seeking to maximize the logarithm of the likelihood (\ref{likelihood}).
The log-likelihood is not concave, and there appears to be no closed form solution to the maximization problem.
Thus one must to rely on iterative algorithms that
produce a sequence of parameters $\theta^{(t)}$ converging to a (local) maximum of the likelihood.
A direct approach employing a generic software routine for numerical maximization of the likelihood function would be computationally intensive,
as each iteration would require multiple (at least $2^K$) calculations of the likelihood function around the estimate obtained at the previous
iteration. A much faster convergence can be achieved by applying a modification of Expectation Maximization (EM)
algorithm.  Details are given below.

We treat the unobserved tissue-specificity information vector
$\bGam_{\lambda}\in\{0,1\}^K$ as a latent variable. The joint likelihood of both observed and latent variables is:
\[
L(\bz,\bgam \, | \, \theta)
\ = \
p_{{\bgam}} \, f_{\bgam} ( \bz \, | \, \theta ) .
\]
The EM algorithm operates in an iterative fashion.
Let $\theta^{(t)} = ( \bmu_0^{(t)}, \Delta^{(t)}, \Sigma^{(t)}, \bp^{(t)})$ be the estimate of the model
parameters after $t$ iterations.
The estimate $\theta^{(t+1)}$ is defined by
\[
\theta^{(t+1)} \ = \ \argmax_\theta Q(\theta : \theta^{(t)}),
\]
where
\[
Q(\theta : \theta^{(t)}) = \sum_\lambda \E_{\bGam_\lambda|\bz_{\lambda},\theta^{(t)}} \big[ \log L(\bz_{\lambda},\bGam_\lambda|\theta) \big] .
\]
The expectation of the log-likelihood is calculated with respect to the conditional distribution
of $\bGam_\lambda$ given the observed vector of correlations $\bz_{\lambda}$ and the model
parameters $\theta^{(t)}$.

Consider the conditional expectation appearing in $Q(\theta : \theta^{(t)})$.  Let $p(\bgam \, | \, \theta)$ denote
the probability of the configuration $\bgam$ under the probability mass function $\bp$ associated with the parameter
$\theta$, and define
\[
p( \bgam \, | \, \bz, \theta)
\ = \
\pr(\bGam_\lambda = \bgam \, | \, \bz, \theta)
\ = \
\frac{ p(\bgam \, | \, \theta) f_{\bgam} ( \bz \, | \, \theta )}
       {\sum_{\bgam'} p(\bgam' \, | \, \theta) f_{\bgam'} ( \bz \, | \, \theta )}
\]
The objective function $Q(\theta : \theta^{(t)})$ then has the form
\[
Q(\theta : \theta^{(t)}) =
\sum_\lambda \sum_{\bgam}
p( \bgam \, | \, \bz_\lambda, \theta^{(t)})
\big[ \log p (\bgam \, | \, \theta) + \log f_{\bgam}( \bz_{\lambda} \, | \, \theta ) \, \big]
\]
Maximization of $Q$ with respect to $\theta$ leads to the explicit formula
\[
p( \bgam \, | \, \theta^{(t+1)}) \ = \  \sum_\lambda p( \bgam \, | \, \bz_\lambda, \theta^{(t)})  \Big/ |\Lambda|
\]
where $|\Lambda|$ is the number of gene-SNP pairs under consideration.
There appears to be no closed form solution for the iterates of $\bmu_0^{(t)}$, $\Sigma^{(t)}$ and $\Delta^{(t)}$.
However, in practice, most of the probability mass of $\bp$ is concentrated at the two
extreme cases $\bgam = \bf 0$ and $\bgam = \bf 1$, reflecting the fact that most transcript-SNP pairs
are associated in no tissues or all tissues.
Approximating $Q(\cdot)$ by restricting  the second sum to $\bgam = 0,1$ leads to explicit (approximate)
estimates of $\bmu_0$, $\Sigma$ and $\Delta$ via the following first order conditions:
\begin{eqnarray*}
	\Delta^{(t+1)} & = & \sum_\lambda p( {\bf 0} \, | \, \bz_\lambda, \theta^{(t)})  \bz_{\lambda}\bz_{\lambda}^T
	\Big/  \sum_\lambda p( {\bf 0} \, | \, \bz_\lambda, \theta^{(t)}) \\[.1in]
	\bmu_0^{(t+1)} & = & \sum_\lambda p( {\bf 1} \, | \, \bz_\lambda, \theta^{(t)}) \bz_{\lambda} \Big/  \sum_\lambda p( {\bf 1} \, | \, \bz_\lambda, \theta^{(t)})\\[.1in]
	\Sigma^{(t+1)} + \Delta^{(t+1)}
	& = &
	\sum_\lambda p( {\bf 1} \, | \, \bz_\lambda, \theta^{(t)}) (\bz_{\lambda} - \bmu_0^{(t+1)})(\bz_{\lambda} - \bmu_0^{(t+1)})^T
	\Big/  \sum_\lambda p( {\bf 1} \, | \, \bz_\lambda, \theta^{(t)})
\end{eqnarray*}
At some iterations the estimates $\Sigma^{(t+1)}$ may fail to be non-negative definite.
In such cases we force $\Sigma^{(t+1)}$ to be non-negative definite by calculating its singular value decomposition
and dropping terms with negative coefficients (negative eigenvalues).

Starting with an initial parameter value $\theta^{(0)}$, we perform sequential updates in the manner described above
until the change in the likelihood falls below a pre-set threshold.
To assess the reliability of the estimate one may run the algorithm multiple times using distinct starting points.
In our experiments the algorithm tends to converge to the same estimate regardless of the starting point.

\section{Proof of Lemma 2.1}\label{lemprf}
\begin{proof}
Let $S$ be a subset of $\{1,\ldots, K\}$ with cardinality $|S| = r$.
It follows from the defining properties of the multivariate normal
distribution that if ${\bf U} \sim {\cal N}_K (\bmu, A)$ then
${\bf U}_{\scriptscriptstyle S} \sim {\cal N}_r (\bmu_{\scriptscriptstyle S}, A_{\scriptscriptstyle S})$.
Therefore we have that
\be\label{zs}
\bZ_{\scriptscriptstyle S}
\ \sim \
\sum_{\bgam \in \{0,1\}^K} p_{\bgam} \,
{\cal N}_r \big( (  \bmu_{0} \cdot \bgam)_{\scriptscriptstyle S}, \,
(\Delta +   \Sigma \cdot \bgam \bgam^T)_{\scriptscriptstyle S} \big)
\ee
Here and in the remainder of the proof we follow the convention that $\bgam$ ranges over $\{0,1\}^K$, and $\bzeta$ ranges over $\{0,1\}^r$.
Elementary arguments show that
\[
(  \bmu_{0} \cdot \bgam)_{\scriptscriptstyle S}
\ = \
  \bmu_{0,\scriptscriptstyle S} \cdot \bgam_{\scriptscriptstyle S}
\ \mbox{ and } \
(\Delta +  \Sigma \cdot \bgam \bgam^T)_{\scriptscriptstyle S}
\ = \
\Delta_{\scriptscriptstyle S} +  \Sigma_{\scriptscriptstyle S} \cdot \bgam_{\scriptscriptstyle S} \bgam_{\scriptscriptstyle S}^T
\]
It then follows from (\ref{zs}) that
\begin{eqnarray*}
\bZ_{\scriptscriptstyle S}
& \sim &
\sum_{\bgam\in \{0,1\}^K} p_{\bgam} \,
{\cal N}_r
\big(  \bmu_{0,\scriptscriptstyle S} \cdot \bgam_{\scriptscriptstyle S}, \,
\Delta_{\scriptscriptstyle S} +  \Sigma_{\scriptscriptstyle S} \cdot \bgam_{\scriptscriptstyle S} \bgam_{\scriptscriptstyle S}^T
\big) \\[.1in]
& = &
\sum_{\bzeta \in \{0,1\}^r} \,
\sum_{\bgam: \bgam_{\scriptscriptstyle S} = \bzeta}
p_{\bgam} \,
{\cal N}_r
\big(  \bmu_{0,\scriptscriptstyle S} \cdot \bgam_{\scriptscriptstyle S}, \,
\Delta_{\scriptscriptstyle S} +\Sigma_{\scriptscriptstyle S} \cdot \bgam_{\scriptscriptstyle S} \bgam_{\scriptscriptstyle S}^T
\big) \\[.15in]
& = &
\sum_{\bzeta \in \{0,1\}^r} \,
{\cal N}_r
\big( \bmu_{0,\scriptscriptstyle S} \cdot \bzeta , \,
\Delta_{\scriptscriptstyle S} + \Sigma_{\scriptscriptstyle S} \cdot \bzeta \bzeta^T
\big)
\sum_{\bgam: \bgam_{\scriptscriptstyle S} = \bzeta}  p_{\bgam} \\[.15in]
& = &
\sum_{\bzeta \in \{0,1\}^r}p_{\bzeta,\scriptscriptstyle S} \,
{\cal N}_r \big( \bmu_{0,\scriptscriptstyle S} \cdot \bzeta , \,
\Delta_{\scriptscriptstyle S} + \Sigma_{\scriptscriptstyle S} \cdot \bzeta \bzeta^T
\big) ,
\end{eqnarray*}
which is the desired expression for distribution of $\bZ_{\scriptscriptstyle S}$.
\end{proof}

\section{Proof of Theorem 3.2}
\label{FDRProof}

\subsection{Continuity and Monotonicity of $F(t)$}

\begin{lem}
\label{lem:gu}
Let $U$ be a bounded, non-negative random variable.  For $t \geq 0$ define
\be
\label{eqn:defg}
G(t)
\ = \
\E[ \, U \, | \, U \! \leq t \, ]
\ = \
\frac{\E[ \, U \, \In( U \! \leq t) \, ]}{ \pr( U \! \leq t )} .
\ee
Then the following hold:
\begin{enumerate}

\item
\label{it:G1}
$G$ is non-decreasing and right continuous;

\item
\label{it:G2}
If $\pr(U = t) = 0$ then $G$ is continuous at $t$;

\item
\label{it:G3}
If $\pr(a < U < b) > 0$ for each $0 < a < b < L$ then
$G$ is strictly increasing on $(0,L)$.

\end{enumerate}

\end{lem}

\noindent
\begin{proof}
To show that $G$ is non-decreasing it suffices to show that
$G(t + \delta) - G(t) \geq 0$ for each fixed $t \geq 0$ and $\delta > 0$.
If $G(t) = 0$ then the result is immediate as the function $G$ is non-negative.
If $G(t)$ is positive, then
\begin{eqnarray*}
G(t + \delta) - G(t)
& = &
\frac{\E[ \, U \, \In( U \! \leq t + \delta) \, ]}{ \pr( U \! \leq t + \delta)}
\ - \
\frac{\E[ \, U \, \In( U \! \leq t) \, ]}{ \pr( U \! \leq t )} \\[.1in]
& = &
\frac{ \E[ \, U \, \In( U \! \leq t + \delta) \, ] \, \pr( U \! \leq t ) \ - \ \E[ \, U \, \In( U \! \leq t) \, ] \, \pr( U \! \leq t + \delta)}
       {\pr( U \! \leq t + \delta) \, \pr( U \! \leq t)} .
\end{eqnarray*}
By elementary arguments the numerator of the last fraction can be expressed as
\begin{eqnarray}
\label{EUT}
\lefteqn{\E[ \, U \, \In( t < U \! \leq t + \delta) \, ] \, \pr( U \! \leq t )
\ - \
\E[ \, U \, \In( U \! \leq t) \, ] \, \pr( t < U \! \leq t + \delta)} \nonumber \\[.05in]
& \geq &
\label{eqn:tp}
t \, \pr( t < U \! \leq t + \delta) \, \pr( U \! \leq t ) \ - \ t \, \pr( U \! \leq t ) \, \pr( t < U \! \leq t + \delta) \\[.05in]
& = &
0 \nonumber .
\end{eqnarray}
Thus $G$ is non-decreasing.  Right continuity of $G$ follows by applying the monotone convergence theorem to the
numerator and denominator in (\ref{eqn:defg}).
If $\pr(U = t) = 0$ then continuity of $G$ at $t$ follows from the dominated convergence theorem
in a similar fashion.  Finally, if $\pr( t < U \! < t + \delta) > 0$ then the inequality in (\ref{eqn:tp}) is strict,
and the final claim follows by considering $t \in [0,L)$ and $\delta > 0$ such that $t + \delta < L$.
\end{proof}

\vskip.2in

\begin{lem}
\label{GLD}
For $i = 0,\ldots,m$ let $f_i$ be the density of the $d$-variate normal distribution
${\cal N}_d(\mu_i, \Sigma_i)$
and let $c_1,\ldots, c_m$ be positive constants.
If at least one of $f_1,\ldots, f_m$ is not equal to $f_0$, then
\[
m_d \Big( \Big\{ x : f_0(x) = \mbox{$\sum_{j=1}^m$} \, c_j \, f_j(x) \Big\} \Big) = 0
\]
where $m_d(\cdot)$ denotes Lebesgue measure on $\real^d$.
\end{lem}

\noindent
\begin{proof}
Define $h(x) = f_0(x) - \sum_{j=1}^m c_j \, f_j(x)$ and let $A = \{ x : h(x) = 0 \}$.
As $h$ is continuous, $A$ is a closed subset of $\real^d$.
We establish the result by way
of contradiction.
Consider first the case in which $d = 1$ and $h(x) = 0$ for each $x \in \real$.
By an easy argument, we can assume that the densities $f_i$, $i = 0,1,\ldots, m$ are distinct
and that $m \geq 1$.
Let $\mu_i$ and $\sigma_i$ be, respectively, the mean and
variance of the distribution specified by the density $f_i$.  Let $(\sigma_j, \mu_j)$ be the
largest element, under the usual lexicographic order, of the set $\{ (\sigma_i, \mu_i) : 0 \leq i \leq m \}$.
Considering the limit of $h(x) / f_j(x)$ as $x$ tends to infinity, we conclude that $c_j = 0$ if $j \neq 0$ or $1 = 0$ if
$j = 0$.  In either case we obtain a contradiction, and therefore $h(x)$ cannot be identically equal to zero.

The remainder of the proof proceeds by induction on $d$.
Consider first the case $d=1$. Note that $h(x)$ is an analytic function of the real variable $x$.
If $m_1(A) > 0$ then there exists $M < \infty$ such that
$m_1(A \cap [-M,M]) > 0$.  In particular, there are infinitely many points of $A$ in the
compact set $[-M,M]$.  Thus $A$ has a limit point $x_0$, and $h(x_0) = 0$ as $A$ is closed.
As the zeros of a non-zero analytic function are necessarily isolated, it follows that $h(x)$ is
identically zero.  This contradicts the argument given above, and we conclude that $m_1(A) = 0$.

Assume now that the lemma holds for dimensions $1, \ldots, d-1$, and consider the
general case of dimension $d$.  Suppose that $m_d(A) > 0$.
By Fubini's theorem, there exist a Borel measurable
set $B \subset \real$ such that (i) $m_1(B) > 0$ and (ii) for every $x_d \in B$ the
section
\[
A(x_d) = \{ x_1^{d-1} : (x_1^{d-1}, x_d) \in A \} \subseteq \real^{d-1}
\]
has $(d-1)$-dimensional Lebesgue measure greater than zero.  (Here $x_1^{d-1}$ denotes the
ordered sequence $x_1, \ldots, x_{d-1}$.)
Note that $h(x) = 0$ can be written in the equivalent form
\be
\label{zero}
0
\ = \
f_0(x_1^{d-1} \, | \, x_d) \, f_0(x_d) \ - \ \sum_{j=1}^m c_j \, f_j(x_1^{d-1} \, | \, x_d) \, f_j(x_d)
\ \ \ \ x \in A
\ee
where $f_j(x_1^{d-1} \, | \, x_d)$ denotes the conditional density of $x_1^{d-1}$ given $x_d$ under
$f_j$, and $f_j(x_d)$ denotes the marginal density of $x_d$ under $f_j$.
If for each $x_d \in B$ the conditional densities $f_j(x_1^{d-1} \, | \, x_d)$ are equal
on $A(x_d)$ then (\ref{zero}) becomes
\[
0 \ = \ f_0(x_d) \ - \ \sum_{j=1}^m c_j \, f_j(x_d)
\ \ \ \ x_d \in B,
\]
which contradicts the induction hypothesis.  Suppose then that for some $x_d \in B$ the conditional densities
$f_j(x_1^{d-1} \, | \, x_d)$ are not all equal on $A(x_d)$.  Then equation (\ref{zero}) becomes
\[
0
\ = \
f_0(x_1^{d-1} \, | \, x_d) \ - \ \sum_{j=1}^m c_j' \, f_j(x_1^{d-1} \, | \, x_d)
\ \ \ \ x_1^{d-1} \in A(x_d)
\]
where $c_j' = c_j \, f_j(x_d) / f_0(x_d)$.   Our assumption regarding the conditional densities ensures that
$f_j(x_1^{d-1} \, | \, x_d)$ is different from $f_0(x_1^{d-1} \, | \, x_d)$ for some $j \geq 1$, again contradicting
the induction hypothesis.   This completes the proof.
\end{proof}

\vskip.2in

\begin{lem}
\label{lem:etaprop}
Let $\eta(\bz)$ be local false discovery rate defined as
\bes
\eta(\bz)
\ : = \
\pr( \bGam = {\bf 0} \, | \, \bZ = \bz)
\ = \
\frac{ p_{\bf 0} f_{\bf 0} (\bz) }{ f(\bz) } .
\ees
and assume that every
diagonal entry of $\Sigma$ is positive.  Then the following hold.
\begin{enumerate}

\item
\label{eta-1}
$\inf_{\bz \in \real^d} \eta(\bz) = 0$.

\item
\label{eta-2}
For every $c \geq 0$ the Lebesgue measure of the set $\{ \bz : \eta(\bz) = c \}$ in $\real^K$ is zero.

\end{enumerate}
\end{lem}

\noindent
\begin{proof}
{\bf Proof of \ref{eta-1}:} As $\eta(z)$ is always positive, it is enough to show that there exists
$\bz \in \real^d$ and $\bgam \in \{0,1\}^K$ such that $f_{\bf 0} (b \bz) / f_{\bgam}(b \bz) \to 0$
as $b \to \infty$.  From the exponential form of the multivariate normal densities,
it can be seen that the last relation will hold if the matrix
$\Delta^{-1} - (\Delta +  \Sigma \cdot \bgam \bgam^T)^{-1}$
has an eigenvalue greater than zero.

Let $\bx_0$ be an eigenvector of the matrix $\Delta$ corresponding
to the smallest eigenvalue $\lambda_{\min}(\Delta)$ (which is positive by assumption).
Assume without loss of generality that $|| \bx_0 || = 1$.
Using the variational formula for eigenvalues, and the relationship between
the eigenvalues of a matrix and those of its inverse, we find that
\begin{eqnarray*}
\lambda_{\max} (\Delta^{-1} - (\Delta +  \Sigma \cdot \bgam \bgam^T)^{-1})
& = &
\max_{z : ||z|| = 1} z^T (\Delta^{-1} - (\Delta +  \Sigma \cdot \bgam \bgam^T)^{-1}) z \\
& \geq &
\max_{z : ||z|| = 1} z^T \Delta^{-1} z
\ - \
\max_{z : ||z|| = 1} z^T (\Delta +  \Sigma \cdot \bgam \bgam^T)^{-1} z \\
& = &
\lambda_{\max} (\Delta^{-1}) \ - \ \lambda_{\max} ((\Delta +  \Sigma \cdot \bgam \bgam^T)^{-1}) \\
& = &
\lambda_{\min} (\Delta) \ - \ \lambda_{\min} (\Delta +  \Sigma \cdot \bgam \bgam^T) \\
& \geq &
\bx_0^T \Delta \bx_0 \ - \  \bx_0^T (\Delta +  \Sigma \cdot \bgam \bgam^T) \bx_0 \\
& = &
\bx_0^T (\Sigma \cdot \bgam \bgam^T) \bx_0
\end{eqnarray*}
Let $1 \leq i \leq K$ be any index for which $x_{0,i} \neq 0$.
If $\bgam$ is the binary $K$-vector having a $1$ in position $i$ and all other
entries equal to $0$, then it is easy to see that the last expression above is
$\sigma_{ii} \, x_{0,i}^2$, which is positive.

\vskip.1in

\noindent
{\bf Proof of \ref{eta-2}:} This follows immediately from Lemma \ref{GLD}

\vskip.2in

\begin{prop}
\label{prop:FIM}
The function $F(t)$ defined as
\bes
F(t)
\ := \
\E ( \eta(\bZ) \, | \, \eta(\bZ) \leq t)
\ = \
\frac{\E[ \eta(\bZ) \, \In(\eta(\bZ) \leq t) ]}{\pr(\eta(\bZ) \leq t)} .
\ees
is continuous and strictly increasing on the interval $(0,L_\eta)$,
where $L_\eta = \sup_{\bz \in \real^d} \eta(\bz) < 1$.
\end{prop}

\noindent
{\bf Proof:} Note that $F(t)$ is of the form $g(t)$ in (\ref{eqn:defg}) with $U = \eta(\bZ)$.  Part
\ref{eta-2} of Lemma \ref{lem:etaprop} establishes that $\pr(\eta(bZ) = t ) = 0$, and continuity of $F$ then
follows from Lemma \ref{lem:gu}.  For $0 < a < b < L_\eta$ we have
\[
\pr(a < \eta(\bZ) < b) \ = \ \pr( \eta(\bZ) \in (a,b) ) \ = \ \pr( \bZ \in \eta^{-1}(a,b) ) .
\]
As $\eta(\bz)$ is continuous $\eta^{-1}(a,b)$ is an open subset of $\real^d$.  Moreover,
$\eta^{-1}(a,b)$ is non-empty by Part \ref{eta-1} of Lemma \ref{lem:etaprop}.
Thus $\pr(a < \eta(\bZ) < b) > 0$ as the density $f$ of $\bZ$ is positive on $\real^d$.  Continuity of $F(t)$
then follows from Lemma \ref{lem:gu}.
\end{proof}

\subsection{Proof of Theorem 3.2}

\begin{lem}
\label{gquant}
Let $G_1, G_2,\ldots : [0,1] \to \real$ be non-decreasing functions.
For fixed $\alpha \in (0, L_\eta)$ define
$\theta_n = \sup\{ t : G_n(t) \leq \alpha \}$
and let $\theta \in (0,1)$ be the unique number such that $F(\theta) = \alpha$.
If $G_n(t) \to F(t)$ for each $t$ in a dense subset $T$ of $[0,1]$ then $\theta_n \to \theta$.
\end{lem}

\noindent
\begin{proof}
Suppose by way
of contradiction that $|\theta_n - \theta| \not\to 0$.  Then there exists $\delta_1, \delta_2 > 0$
such that $\{ \theta - \delta_1, \theta + \delta_2 \} \subseteq T$ and an
infinite subsequence $n_k$ of $1, 2, \ldots$ such that
either $\theta_{n_k} \leq \theta - 2 \delta_1$ for each $k \geq 1$ or
$\theta_{n_k} \geq \theta + 2 \delta_2$ for each $k \geq 1$.  In the first case, the definition of $\theta_n$
and the monotonicity of $G_n$ imply
\[
\alpha \ \leq \ G_{n_k}(\theta_{n_k} + \delta_1) \ \leq \ G_{n_k}(\theta - \delta_1)
\]
Taking limits as $k \to \infty$ we find
$
\alpha \ \leq \ F(\theta - \delta_1) < \alpha
$
as $F$ is strictly increasing, which is a contradiction.  In the second case, a similar argument shows that
\[
\alpha \ \geq \ G_{n_k}(\theta_{n_k} - \delta_2) \ \geq \ G_{n_k}(\theta + \delta_2) .
\]
Taking limits as $k \to \infty$ yields
$
\alpha \ \geq \ F(\theta + \delta_2) > \alpha,
$
which is again a contradiction.  This concludes the proof.
\end{proof}

\vskip.2in

\noindent
{\bf Proof of Theorem 3.2:} 
Let $\hat{\theta}_n = \sup\{ t : \hat{F}_n(t) \leq \alpha \}$ and let $\theta$ be the unique number such that
$F(\theta) = \alpha$.
We claim that $\hat{\theta}_n \to \theta$ in probability.  To show this, assume to the contrary that
there exists $\delta > 0$ and a subsequence
$n_k$ such that
\be
\label{deltsep}
\pr\big( | \hat{\theta}_{n_k} - \theta | > \delta \big) > \delta \ \ \mbox{ for each } \ \ k \geq 1.
\ee
Let $T$ be any countable, dense subset of $[0,1]$.  Our assumptions imply that $\hat{F}_n(t) \to F(t)$
in probability for each $t \in T$.  By a standard diagonalization argument, there exists a subsequence
$m_k$ of $n_k$ such that $\hat{F}_{m_k}(t) \to F(t)$ with probability one for each $t \in T$.  It then follows
from
Lemma \ref{gquant} that $\hat{\theta}_{m_k} \to \theta$ with probability one, which contradicts (\ref{deltsep}).

In order to establish the theorem, it will be convenient to work with version of $M_n$ and $N_n$
in which the data-dependent threshold $\hat{\theta}_n$ is replaced by the limiting value $\theta$.
Define
\[
\tilde{M}_n \ = \ \sum_{\lambda \in \Lambda_n} \In(\bGam_\lambda = 0) \,
              \In(\eta(\bZ_\lambda) \leq \theta )
\ \ \mbox{ and } \ \
\tilde{N}_n \ = \  \sum_{\lambda \in \Lambda_n}
              \In(\eta(\bZ_\lambda) \leq \theta )
\]
Note that $\E \tilde{N}_n = |\Lambda_n| \cdot \pr( \eta(\bZ) \leq \theta)$.
By an elementary conditioning argument,
\begin{eqnarray*}
\E \tilde{M}_n
& = &
\sum_{\lambda \in \Lambda_n}
\E \Big\{ \pr( \bGam_\lambda = 0 \, | \, \bZ_\lambda ) \,
              \In(\eta(\bZ_\lambda) \leq t_n(\alpha) ) \Big\}  \\
& = &
\sum_{\lambda \in \Lambda_n}
\E \Big\{ \eta( \bZ_\lambda ) \,
              \In(\eta(\bZ_\lambda) \leq t_n(\alpha) ) \Big\}  \\[.1in]
& = &
|\Lambda_n| \cdot \E[ \eta(\bZ) \, \In( \eta(\bZ) \leq t ) ]  .
\end{eqnarray*}
For each $\delta > 0$,
\begin{eqnarray*}
\E | \tilde{N}_n - N_n |
& \leq &
\sum_{\lambda \in \Lambda_n}  \pr (\eta(\bZ_\lambda) \in [\hat{\theta}_n, \theta] \cup [\theta, \hat{\theta}_n] ) \\[.1in]
& \leq &
|\Lambda_n| \, \big[ \pr \big( \eta(\bZ) \in (\theta - \delta, \theta + \delta) \big)
               + \pr \big( | \hat{\theta}_n - \theta | \geq \delta \big) \big] .
\end{eqnarray*}
As $\hat{\theta}_n \to \theta$ in probability and the distribution of $\eta(\bZ)$ has no point masses, the
last inequality implies that $\E | \tilde{N}_n - N_n | = |\Lambda_n| \cdot o(1)$.
A similar argument shows that $\E | \tilde{M}_n - M_n | = |\Lambda_n| \cdot o(1)$.  Thus as $n$ tends to
infinity,
\begin{eqnarray*}
\frac{\E M_n}{\E N_n}
& = &
\frac{\E \tilde{M}_n + |\Lambda_n| \cdot o(1)}{\E \tilde{N}_n + |\Lambda_n| \cdot o(1)} \\[.1in]
& = &
\frac{\E[ \eta(\bZ) \, \In( \eta(\bZ) \leq \theta ) ] + o(1)}{\pr( \eta(\bZ) \leq \theta) + o(1)} \\[.1in]
& \to &
\frac{\E[ \eta(\bZ) \, \In( \eta(\bZ) \leq \theta ) ]}{\pr( \eta(\bZ) \leq \theta)}
\ = \
F(\theta)
\ = \
\alpha .
\end{eqnarray*}
This completes the proof of the theorem.

\section{Simulation Study}
\label{Sims}

In this section, we examine the performance of MT-eQTL through a simulation study with $K = 4$ tissues.
As the basis of the model and subsequent inferences is the collection of z-statistic vectors
derived from the observed genotype and transcript data, we directly simulate the z-statistics.

\subsection{Simulation Setting}
We simulate 10 million vectors $\bz_\lambda$ independently from the MT-eQTL model using
parameters $\theta = (\Delta, \Sigma, \bp,\bmu_0)$ obtained from eQTL analysis of data from the GTEx initiative (we consider the tissues blood, lung, muscle, and thyroid, which we denote by a, b, c, and d, respectively).
Sample sizes, sample overlap, and degrees of freedom after covariate correction are given in Table \ref{tab:df}.
The true model parameters are given in Table \ref{tab:param}.
Note that the average effect size parameter $\bmu_0$ is set to be zero in data generation and model fitting for simplicity.
We remark that allowing $\bmu_0$ to be free has little effect on the numerical results.

We simulated each vector $\bz_\lambda$ in a two-step fashion: first drawing $\bgam \in \{0,1\}^4$ from $\bp$, and then
drawing $\bz_\lambda$ from $f_{\bgam}(\bz)$ given $\bgam$.  Access to the true configurations $\bgam$ enables us to
assess false discovery rates associated with inferences from the fitted model.

\subsection{Model Fit}

The approximate EM procedure was used to fit the full 4-dimensional model, as well as all possible 1-, 2-, and 3-dimensional models.
We terminated EM updates when the difference between log likelihoods in two consecutive iterations was less than $0.01$.
The average number of iterations until convergence of the EM procedure was 80.
The running time of the EM procedure depended on the number of tissues in the model, ranging from
about 1 second per iteration for the 1-dimensional models to about 40 seconds per iteration for full 4-dimensional model on a standard desktop PC.
Fitting of the 4-dimensional model based on the simulated data took slightly more than one hour.

As expected, the parameters estimated from the simulated data are very close to those
used to generate the data.  For the 4-dimensional model, the relative error
of each entry of $\Sigma$ is less than $0.3\%$, while the relative error for each entry of $\Delta$
is less than $0.7\%$.  For the probability mass vector $\bp$, thirteen of sixteen entries had relative error
less than $1\%$, with the remaining relative errors equal to $1.45\%$, $1.66\%$ and $4.31\%$.
These results confirm that the approximate EM procedure works well on the simulated data.

\subsection{Results}

{
We apply the adaptive thresholding procedure to the local FDRs and detect eQTLs with different configurations.
In particular, we attempt to identify eQTLs in at least one tissue,
in a single tissue, and in all tissues. 
The corresponding null configurations are shown in the second column of Table \ref{tab:MAP}.
In all studies, we fixed the nominal FDR threshold at $\alpha=0.05$.
Table \ref{tab:MAP} contains the number of true alternative cases, the number of total discoveries,
the number of overlaps (i.e., true positives), the true positive rate (TPR), and the FDR
in each study.
}

{In all studies,
the observed FDRs are strictly below the nominal level of $5\%$; and the TPRs are around $40\%$.
These TPRs are considered relatively high because many alternative cases may have modest to small effect sizes
in the simulated data, and are not readily distinguishable from the null cases.
This behavior is representative of real data where signals are not always highly identifiable.
{\color{black} The TPR for testing cross-tissue eQTL is the lowest among all cases because it is the most challenging problem: a cross-tissue eQTL may be easily mistaken as other eQTL configurations (e.g., 1110, 1101, 1011, 0111, etc). Nonetheless, the TPR for such case is still reasonably high, which demonstrates the efficacy of the proposed method.
In addition, we emphasize that our method is flexible enough to detect eQTL of different configurations.
Numerical results indicate that the method has great power in identifying significant association between a gene and a SNP in a single tissue or in any tissue. }}


In order to assess how the use of auxiliary tissues increases statistical power of detecting eQTL in a target tissue,
we fit a series of nested MT-eQTL models for tissue sets $\{$a$\}$, $\{$a,b$\}$, $\{$a,b,c$\}$, and $\{$a,b,c,d$\}$ and only focused
on eQTL detection in tissue a.  In each study, we fixed the target set $S=\{$a$\}$, and applied the adaptive thresholding procedure to the marginal local FDR defined in Section 3.3 in the main article.
We set the nominal FDR at the level of $0.05$ for all studies.
As a result, the discoveries from different studies are comparable, as they are all detected gene-SNP pairs with eQTL in tissue a at the FDR of $0.05$.
Table \ref{tab:gain} shows the TPRs and FDRs in different studies.
The number of true alternative cases is 1,596,410.
The TPRs increase steadily with the number of auxiliary tissues considered in the analysis, while the FDRs are all controlled at the nominal level.
The result indicates that by borrowing information from auxiliary tissues, the model gains power of detecting eQTL in a target tissue without inflating the FDR.
Similar results hold for more sophisticated hypothesis testings. 

\section{GTEx Estimations}
\label{sec:GTEx est}

{\color{black}The sample information of the GTEx pilot data is provided in Figure \ref{GTExSample}.}
The estimated model parameters $\Delta$ and $\Sigma$ for the GTEx data are given below.
The tissues are ordered alphabetically.
The parameter $\bmu_0$ was set to zero.
The estimated mass function $\bp$ (prior probabilities for 512 configurations) is provided in a separate text file (SuppC-p.txt) due to space limitations.

\begin{figure}[htbp]
\begin{centering}
\includegraphics[width=3.6in]{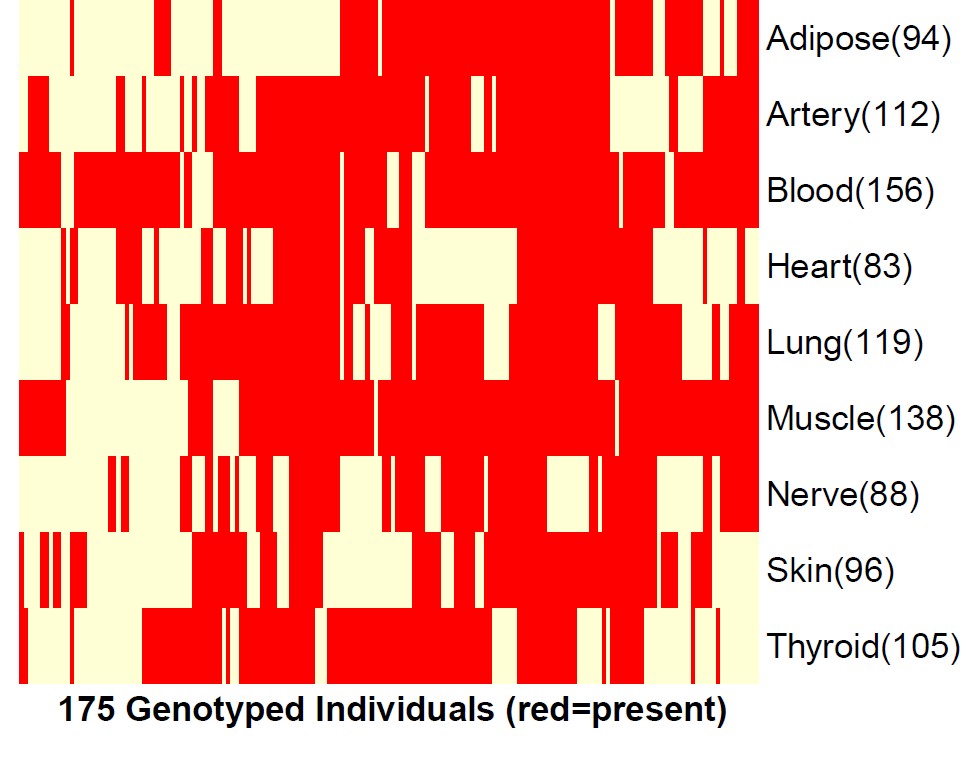}
\caption{Sample information of the GTEx data.
Each column represents a genotyped individual with expression measurements in at least one tissue; each row corresponds to a tissue. Red means the individual is a donor of the corresponding tissue. }
\label{GTExSample}
\end{centering}
\end{figure}

\vskip.1in
$
\Delta=\begin{pmatrix}
    1.0000  &  0.1704  &  0.0923  &  0.1010  &  0.1390    &0.1409 &   0.1687   & 0.1415  &  0.1441\\
    0.1704  &  1.0000   & 0.0960 &   0.1179   & 0.1518   & 0.1460  &  0.1942   & 0.1336  &  0.1491\\
    0.0923  &  0.0960   & 1.0000    &0.0779    &0.1312    &0.0780   & 0.1007   & 0.0890  &  0.1032\\
    0.1010  &  0.1179   & 0.0779   & 1.0000&    0.1268   & 0.1192    & 0.1093   & 0.0893  &  0.1247\\
    0.1390  &  0.1518   & 0.1312  &  0.1268 &   1.0000  &  0.1188    & 0.1543   & 0.1220  &  0.1767\\
    0.1409  &  0.1460  &  0.0780 &   0.1192  &  0.1188 &   1.0000    & 0.1366   & 0.1095  &  0.1258\\
    0.1687  &  0.1942   & 0.1007    &0.1093   & 0.1543    &0.1366    & 1.0000   & 0.1372  &  0.1477\\
    0.1415  &  0.1336 &   0.0890   & 0.0893    &0.1220   & 0.1095    & 0.1372   & 1.0000  &  0.1097\\
    0.1441  &  0.1491  &  0.1032  &  0.1247   & 0.1767  &  0.1258    & 0.1477  &  0.1097  &  1.0000
    \end{pmatrix}$,

$
\Sigma=\begin{pmatrix}
    4.2692  &  4.5320    &4.1062&    3.2993  &  4.6078 &   4.0864    & 4.2076  &  3.9694  &  4.4595\\
    4.5320  &  5.4178   & 4.4545 &   3.6526 &   5.0411  &  4.5731    & 4.6975  &  4.3167  &  5.0072\\
    4.1062  &  4.4545  &  6.1588  &  3.3196 &   5.0385   & 4.2452    & 4.0646  &  4.0090  &  4.5213\\
    3.2993  &  3.6526    &3.3196   & 3.2123    &3.7223 &   3.6852    & 3.3418  &  3.1225  &   3.7332\\
    4.6078  &  5.0411   & 5.0385    &3.7223    &5.5488  &  4.5088    & 4.6816  &  4.5263  &  5.2369\\
    4.0864  &  4.5731  &  4.2452&    3.6852   & 4.5088   & 5.1569   & 4.0399  &  3.9304  &  4.3674\\
    4.2076  &  4.6975    &4.0646 &   3.3418  &  4.6816    &4.0399  &  4.5993  &  4.0265  &  4.6699\\
    3.9694  &  4.3167   & 4.0090 &   3.1225 &   4.5263   & 3.9304 &   4.0265  &  4.3420  &  4.4163\\
    4.4595  &  5.0072  &  4.5213  &  3.7332&    5.2369   & 4.3674&    4.6699  &  4.4163  &  5.6492
    \end{pmatrix}$ .

{\color{black}
The fitting times of MT-eQTL and the Meta-Tissue method \citep{sul2013effectively} on a sequence of sub-models of different dimensions based on alphabetically ordered tissues are presented in Table \ref{tab:time}.
(We note that fitting sub-models of MT-eQTL is unnecessary in practice, as one can obtain them through
marginalization of the full model.)
The use of configuration vectors in MT-eQTL makes analysis results more interpretable, but it also makes the runtime of MT-eQTL sensitive to the number of tissues.
Nevertheless, our method is computationally efficient when the number of tissues is moderate.
On the other hand, the Meta-Tissue method is less restricted by the number of tissues, but it is more sensitive to the total number of gene-SNP pairs.
The runtime for Meta-Tissue may quickly become impractical when there are too many gene-SNP pairs.}

\begin{table}[htbp]
\caption{Sample sizes (diagonal), sample overlap (off-diagonal), and degrees of freedom for different tissues in the simulation.}
\label{tab:df}
\small
\begin{center}
\begin{tabular}{|l|c|c|c|c||c|}
\hline
& a  & b & c & d & Degree of Freedom \\
\hline
a & 156 & 104 & 122 & 90 & 137\\
\hline
b & & 119   & 100 & 84 & 100\\
\hline
c & & & 138   & 88 & 119\\
\hline
d & & & & 105 & 86\\
\hline
\end{tabular}
\end{center}
\end{table}

\begin{table}[htbp]
\caption{The true generating model parameters $(\Delta,\Sigma,\bp)$ for the simulation study. The prior probabilities are provided for all possible eQTL configurations represented by 4-digit 0/1 sequences: 0 means no eQTL and 1 indicates the presence of eQTL in a tissue. }
\label{tab:param}
\small
\begin{center}
\subfloat[$\Delta$]{
\begin{tabular}{|l|c|c|c|c|}
  \hline
  & a& b& c& d\\
  \hline
a&    1.0000  &  0.1347 &   0.0805 &   0.1089\\
\hline
 b&   0.1347  &  1.0000 &   0.1204 &   0.1794\\
 \hline
c&    0.0805  &  0.1204 &   1.0000 &   0.1288\\
\hline
d&    0.1089  &  0.1794 &   0.1288 &   1.0000\\
\hline
\end{tabular}}
\vskip.2in

\subfloat[$\Sigma$]{
\begin{tabular}{|l|c|c|c|c|}
  \hline
  & a& b& c& d\\
  \hline
a&     6.5699  &  5.3098  &  4.4683  &  4.7126\\
\hline
b&    5.3098  &  5.9752  &  4.7906  &  5.5778\\
\hline
c&    4.4683  &  4.7906  &  5.5263  &  4.6493\\
\hline
d&    4.7126  &  5.5778  &  4.6493  &  6.0178\\
  \hline
\end{tabular}}

\vskip.2in
\subfloat[$\bp$]{
\begin{tabular}{|l|c|c|c|c|c|c|c|c|}
\hline
Config (abcd) & 0000 & 0001  & 0010 & 0011 & 0100 & 0101& 0110 & 0111 \\
\hline
Prior & 0.7721 & 0.0202 & 0.0190 &  0.0037 & 0.0104 & 0.0033 & 0.0010 & 0.0107\\
\hline
Config (abcd) & 1000 & 1001 & 1010 & 1011 & 1100 & 1101 & 1110 & 1111\\
\hline
Prior &  0.0196 &  0.0010 & 0.0008 & 0.0009 & 0.0029 & 0.0085 &    0.0019 & 0.1240\\
\hline
\end{tabular}}
\end{center}
\end{table}

\begin{table}[htbp]
\caption{A variety of eQTL detection inferences with the MT-eQTL model in the 4-tissue simulation study. From top to bottom, we aim to identify eQTLs: 1) in at least one tissue; 2) in tissue a (the null consists of all configurations with 0 in the first position); 3) in tissue b; 4) in tissue c; 5) in tissue d; 6) in all 4 tissues. }
\label{tab:MAP}
\small
\begin{center}
\begin{tabular}{|c|c|r|r|r|r|r|}
\hline
Index & Null Config & $\#$ Alternative Cases  & $\#$ Discoveries  &  $\#$ Overlaps &  TPR &  FDR \\
\hline
1) &0000  & 2,279,307 & 1,038,456 & 987,083 &  .4331 &.0495 \\
\hline
2) & 0*** & 1,596,410 & 679,207 & 645,700 & .4045 & .0493\\
\hline
3) & *0** & 1,626,961 & 746,265 & 709,258 & .4359 & .0496\\
\hline
4) & **0* & 1,618,655 & 663,367 & 630,502 & .3895 & .0495\\
\hline
5) & ***0 & 1,722,789 & 770,722 & 732,600 & .4252 & .0495\\
\hline
6) & all but 1111 & 1,239,630 & 417,867 & 397,341 & .3205 & .0491\\
\hline
\end{tabular}
\end{center}
\end{table}

\begin{table}[htbp]
\caption{The TPRs and FDRs of {\em detecting eQTL in tissue a} using the MT-eQTL model on tissue sets $\{$a$\}$, $\{$a,b$\}$, $\{$a,b,c$\}$, and $\{$a,b,c,d$\}$.}
\label{tab:gain}
\begin{center}
\begin{tabular}{|c|c|c|c|c|}
\hline
 & $\{$a$\}$ & $\{$a,b$\}$  & $\{$a,b,c$\}$  &  $\{$a,b,c,d$\}$\\
\hline
TPR& .2753& .3475& .3806&.4045\\
\hline
FDR& .0500&.0499 &.0496 &.0493\\
\hline
\end{tabular}
\end{center}
\end{table}

\begin{table}[htbp]
\caption{Approximate fitting times for $k$-dimensional MT-eQTL models and Meta-Tissue methods using the GTEx data.}
\label{tab:time}
\begin{center}
\begin{tabular}{|c|c|c|c|c|c|c|c|c|c|}
\hline
Time& $k=1$  & $k=2$ & $k=3$ & $k=4$  & $k=5$ & $k=6$ & $k=7$ & $k=8$ & $k=9$  \\
\hline
MT-eQTL & $<1$ min & 15 min & 30 min & 1 hr & 2.5hr & 6hr & 11hr & 16 hr & 24 hr\\
\hline
Meta-Tissue & 130 min & 165 min & 165 min & 3 hr & 3 hr & 3.25 hr & 3.5 hr & 4 hr  & 5 hr\\
\hline
\end{tabular}
\end{center}
\end{table}


\end{document}